\documentclass[aps,pra,twocolumn,nofootinbib,superscriptaddress]{revtex4}

\usepackage[english]{babel}
\usepackage{amsmath}
\usepackage{amsthm}
\usepackage{amssymb}
\usepackage{booktabs}
\usepackage{color}
\usepackage{hyperref}
\hypersetup{
  colorlinks   = true,  %Colours links instead of ugly boxes
  urlcolor     = blue,  %Colour for external hyperlinks
  linkcolor    = blue,  %Colour of internal links
  citecolor    = red    %Colour of citations
}
\usepackage{natbib}
\usepackage{pgfplots}
\usepackage{subfigure}
\usepackage{url}
\usepackage[normalem]{ulem}

\usepackage{scalerel,stackengine}
\stackMath
\newcommand\reallywidehat[1]{%
\savestack{\tmpbox}{\stretchto{%
  \scaleto{%
    \scalerel*[\widthof{\ensuremath{#1}}]{\kern-.6pt\bigwedge\kern-.6pt}%
    {\rule[-\textheight/2]{1ex}{\textheight}}%WIDTH-LIMITED BIG WEDGE
  }{\textheight}% 
}{0.5ex}}%
\stackon[1pt]{#1}{\tmpbox}%
}
\usepackage{comment}
\usepackage{float}
\usepackage{bm}

\begin{document}

\title{Three-body universality in ultracold $p$-wave resonant mixtures}

\author{P. M. A. \surname{Mestrom}}
%\email{p.m.a.mestrom@tue.nl}
\altaffiliation[Corresponding author: ]{p.m.a.mestrom@tue.nl}
\affiliation{Eindhoven University of Technology, P.~O.~Box 513, 5600 MB Eindhoven, The Netherlands}

\author{V. E. \surname{Colussi}}
\affiliation{Eindhoven University of Technology, P.~O.~Box 513, 5600 MB Eindhoven, The Netherlands}
\affiliation{INO-CNR BEC Center and Dipartimento di Fisica, Universit\`{a} di Trento, 38123 Povo, Italy}

\author{T. \surname{Secker}}
\affiliation{Eindhoven University of Technology, P.~O.~Box 513, 5600 MB Eindhoven, The Netherlands}

\author{J.-L. \surname{Li}}
\affiliation{Eindhoven University of Technology, P.~O.~Box 513, 5600 MB Eindhoven, The Netherlands}

\author{S. J. J. M. F. \surname{Kokkelmans}}
\affiliation{Eindhoven University of Technology, P.~O.~Box 513, 5600 MB Eindhoven, The Netherlands}

\date{\today}

%\pacs{31.15.-p, 34.50.-s, 67.85.-d \del{??????????}}

\begin{abstract}
We study three-body collisions within ultracold mixtures with resonant interspecies $p$-wave interactions. Our results for the three-body effective interaction strength and decay rate are crucial towards understanding the stability and lifetime of these dilute quantum fluids. On resonance, we find that a class of universal scattering pathways emerges, regardless of the details of the short-range interactions. This gives rise quite generally to a remarkable regime where three-body effective interactions dominate over both inelastic decay and two-body effective interactions. Additionally, we find a series of mass-ratio-dependent trimer resonances further from resonance.
\end{abstract}

\maketitle

\textit{Introduction.}---The physics of $p$-wave interactions is fundamental to many important quantum systems such as superfluid ${}^{3}$He \cite{leggett1975helium3}, unconventional superconductors \cite{kallin2012superconductorSr2RuO4}, polarons \cite{levinsen2012pwavepolaron}, and halo nuclei \cite{ji2016halonuclei3body,hammer2017halonucleiEFT}. This subject has received a recent surge of attention in ultracold atomic gases due to the availability of $p$-wave Feshbach resonances via which the interaction strength can be tuned in both fermionic \cite{regal2003pwaveFR40K,zhang2003pwaveFR6Li,ticknor2004pwaveFR,ahmedbraun2021pwaveFR} and mixed systems \cite{dong2016pwaveFRRb85Rb87,cui2018broadFR}. 
%\textit{Ultracold atoms provide a highly controllable platform to experimentally investigate such quantum systems due to the availability of $p$-wave Feshbach resonances via which the interaction strength can be tuned in both fermionic \cite{regal2003pwaveFR40K,zhang2003pwaveFR6Li,ticknor2004pwaveFR,ahmedbraun2021pwaveFR} and mixed systems \cite{dong2016pwaveFRRb85Rb87,cui2018broadFR}.} 
Recently, a powerful set of universal relations connecting thermodynamical and microscopic properties were found for ultracold Fermi gases with strong $p$-wave interactions \cite{yoshida2015pwaveFermigasContact,yu2015pwaveFermigasContact,luciuk2016pwaveFermigas}, and such systems are predicted to display topological quantum phase transitions \cite{read2000pairedfermions2D,gurarie2005phasetransitionsPwave}. For $p$-wave resonant mixtures, an intriguing finite-momentum atomic-molecular superfluid phase is predicted \cite{radzihovsky2009pwaveResonantBoseMixture, radzihovsky2011pwaveResonantBoseMixture, li2019pwaveResonantBoseMixture}; however, these mixtures remain largely unexplored.

Determining the thermodynamics of ultracold $p$-wave resonant mixtures requires an analysis of microscopic few-body scattering processes. In the case of a mixture of weakly interacting Bose-Einstein condensates (BECs), the miscibility and stability of the system are determined by the intra- and interspecies scattering lengths which set the effective two-body interaction strengths \cite{pethick2002bec,stenger1998spinorBEC,papp2008mixedBECs,mccarron2011mixedBECs,wacker2015mixedBECs,wilson2021collapseMixedBECs}. However, elastic three-body scattering processes can also play a pivotal role through an effective three-body interaction, which was predicted recently to give rise to liquid quantum droplets in single-component BECs at weak interactions \cite{mestrom2020hypervolumeVdW,bulgac2002droplets, zwerger2019phasetransition}. Identifying other regimes dominated by three-body effective interactions and studying the associated evolution from few- to many-body physics remains an important, open pursuit, in particular at strong interactions, which motivates the present study.

Three-body effective interactions are typically ignored in descriptions of ultracold atomic gases due to their diluteness \cite{pitaevskii2016book,borzov2012BECnearFR}. 
%\ins{Just beyond the dilute-gas limit, these interactions are still weak \cite{borzov2012BECnearFR}.} 
In the vicinity of an $s$-wave dimer resonance, these interactions are strong \cite{efimov1979threebody, braaten2002diluteBEC, braaten2006universality, dincao2018review, mestrom2019hypervolumeSqW}, but so are losses \cite{braaten2006universality, naidon2017review, greene2017review, dincao2018review} and resultant heating \cite{makotyn2014unitaryBosegas,eismann2016unitaryBosegas,eigen2017unitaryBosegas,dincao2018unitaryBosegas}. We find that for $p$-wave resonant mixtures this barrier can be overcome quite generally via a set of three-body elastic scattering processes that involve $p$-wave interactions between two dissimilar particles and even occur at zero collision energy. This gives rise to an intriguing regime near a $p$-wave dimer resonance where three-body effective interactions dominate over both losses and two-body effective interactions, which opens the way to novel classes of quantum fluids.

%a wealth of possibilities. 

%Due to the diluteness of ultracold atomic gases, three-body effective interactions are typically negligible compared to their two-body counterparts.  However, near trimer and dimer resonances in single-component Bose gases, the three-body interaction strength is strongly enhanced \cite{efimov1979threebody, braaten2002diluteBEC, braaten2006universality, dincao2018review, mestrom2019hypervolumeSqW}, but so are losses \cite{braaten2006universality, naidon2017review, greene2017review, dincao2018review}. 
%%%%%\ins{In mixtures the situation can be different since scattering processes involving combinations of $s$ and $p$-wave two-body interactions can contribute to the three-body elastic amplitude.} 
%In mixtures the situation can be different due to three-body elastic scattering processes involving $p$-wave interactions between two dissimilar particles that even occur at zero energy.
%%%%%%The presence of $p$-wave interactions between dissimilar particles in mixtures opens up however the possibility of three-body elastic scattering occuring via a combination of $s$ and $p$-wave two-body interactions, even at zero energy when $p$-wave two-body scattering is suppressed. 
%We find that this gives rise to an intriguing regime near a $p$-wave dimer resonance where strong three-body effective interactions dominate over both losses and two-body effective interactions.

In this Letter, we study mixed three-body systems near an interspecies $p$-wave dimer resonance. We extract the elastic transition amplitude for scattering at zero energy, which provides information on both the strength of three-body effective interactions and recombination in ultracold mixtures. We find that this transition amplitude diverges universally on resonance, depending only on a few parameters that characterize low-energy $s$- and $p$-wave two-body collisions. For two identical bosons interacting with a dissimilar particle, we also analyze how a series of trimer resonances, originating from a universal long-range three-body attraction \cite{efremov2013pwaveBBX, zhu2013TwoScatteringCenters}, impacts the elastic three-body transition amplitude near the $p$-wave dimer resonance. %\ins{The number of trimer states depends on the mass ratio of the particles as predicted by Ref.~\cite{efremov2013pwaveBBX} who investigated tri} 
We conclude with a discussion of the experimental and theoretical implications of our findings.

%In this Letter, we study two identical bosons interacting with a dissimilar particle near a $p$-wave dimer resonance. We extract the elastic transition amplitude for scattering at zero energy, which provides information on both the strength of three-body effective interactions and recombination in Bose-Bose and Bose-Fermi mixtures.  We find that this transition amplitude diverges universally on resonance, which is a general feature of three-body scattering of dissimilar particles.  Additionally, we find a series of trimer resonances and study their effect on the elastic three-body transition amplitude.  We conclude with a discussion of the experimental and theoretical implications of our findings.

\textit{Formalism.}---To study three-body scattering, we start from the Alt-Grassberger-Sandhas (AGS) equations \cite{alt1967ags},
%\begin{equation}\label{eq:AGS}
%\begin{aligned}
%U_{0 0}(z) &= \sum_{\alpha = 1}^{3} T_{\alpha}(z) G_0(z) U_{\alpha 0}(z), \\
%U_{\alpha 0}(z) &= G_0^{-1}(z) + \sum_{\substack{\beta = 1 \\ \beta \neq \alpha}}^{3} T_{\beta}(z) G_0(z) U_{\beta 0}(z)
%\\
%\text{ for } \alpha &= 1, 2, 3,
%\end{aligned}
%\end{equation}
\begin{equation}\label{eq:AGS}
\begin{aligned}
U_{\alpha 0}(z) &= (1-\delta_{\alpha 0}) G_0^{-1}(z) + \sum_{\substack{\beta = 1 \\ \beta \neq \alpha}}^{3} T_{\beta}(z) G_0(z) U_{\beta 0}(z)
\\
&\text{ for } \alpha = 0, 1, 2, 3,
\end{aligned}
\end{equation}
%\begin{equation}\label{eq:AGS_v3}
%\begin{aligned}
%U_{\alpha \beta}(z) &= (1-\delta_{\alpha \beta}) G_0^{-1}(z) + \sum_{\substack{\gamma = 1 \\ \gamma \neq \alpha}}^{3} T_{\gamma}(z) G_0(z) U_{\gamma \beta}(z)
%\\
%&\text{ for } \alpha, \beta = 0, 1, 2, 3,
%\end{aligned}
%\end{equation}
which define a set of transition operators $U_{\alpha 0}(z)$ for scattering of three free particles at energy $z$ in their center-of-mass frame. The outgoing states are labeled by $\alpha$ and are either free-particle states ($\alpha = 0$) or a state consisting of a free particle and a dimer, in which case $\alpha = 1, 2, 3$ specifies the free particle. Here, $G_0(z)$ represents the free Green's function $(z - H_0)^{-1}$, where $H_0$ is the three-body kinetic energy operator in the center-of-mass frame. $T_{\alpha}(z)$  describes two-body scattering between particles $\beta$ and $\gamma$ with particle $\alpha$ spectating ($\alpha,\beta,\gamma = 1,2,3$, $\alpha \neq \beta \neq \gamma$). This means that $T_{\alpha}(z) = V_{\beta \gamma} + V_{\beta \gamma} G_0(z) T_{\alpha}(z)$, where $V_{\beta \gamma}$ indicates the pairwise potential between particles $\beta$ and $\gamma$ and is assumed to be spherically symmetric.
%In this study, we consider spherically symmetric potentials.

%\ins{The three-body transition operator $U_{00}(0)$ determines the zero-energy three-body scattering state, where we note that the limit $z \to 0$ is taken always from the upper half of the complex energy plane \cite{SupplMat}.}
The elastic three-body transition operator $U_{00}(z)$ determines the zero-energy three-body scattering state via  $\lvert \Psi_{\mathrm{3b}}(0)\rangle = \lvert \mathbf{0}, \mathbf{0} \rangle + G_0(0) U_{00}(0)\lvert \mathbf{0}, \mathbf{0} \rangle$, when the limit $z \to 0$ is taken from the upper half of the complex energy plane. 
%, where we note that the limit $z \to 0$ is taken always from the upper half of the complex energy plane. %\inss{The state $\lvert \mathbf{0}, \mathbf{0} \rangle$ represents the zero-momentum plane wave state corresponding to the relative motion defined below}. 
Here we also introduce the free-particle states $\lvert \mathbf{p}_{\alpha}, \mathbf{q}_{\alpha} \rangle_{\alpha}$, where $\mathbf{p}_{\alpha} = \mu_{\beta \gamma} (\mathbf{P}_{\beta}/m_{\beta}  - \mathbf{P}_{\gamma}/m_{\gamma})$ and $\mathbf{q}_{\alpha} = \mu_{\beta \gamma,\alpha} [\mathbf{P}_{\alpha}/m_{\alpha} - (\mathbf{P}_{\beta} + \mathbf{P}_{\gamma})/(m_{\beta} + m_{\gamma})]$ are the Jacobi momenta describing the relative motion of the three-particle system and $\mathbf{P}_{\alpha}$ is the lab momentum of particle $\alpha$. The masses $m_{\alpha}$ of particles $\alpha = 1, 2$, and 3 determine the reduced masses
$\mu_{\beta \gamma} = m_{\beta} m_{\gamma}/(m_{\beta} + m_{\gamma})$ and $\mu_{\beta \gamma,\alpha} = m_{\alpha}(m_{\beta} + m_{\gamma})/(m_{\alpha} + m_{\beta} + m_{\gamma})$. 
%%%%%%%%%%where $\mathbf{p}_{\alpha} = \frac{1}{m_{\beta} + m_{\gamma}} (m_{\gamma} \mathbf{P}_{\beta} - m_{\beta} \mathbf{P}_{\gamma})$ and $\mathbf{q}_{\alpha} = \frac{1}{m_{\alpha} + m_{\beta} + m_{\gamma}} [ (m_{\beta} + m_{\gamma}) \mathbf{P}_{\alpha} - m_{\alpha} (\mathbf{P}_{\beta} + \mathbf{P}_{\gamma})]$ are the Jacobi momenta describing the relative motion of the three-particle system and $\mathbf{P}_{\alpha}$ is the lab momentum of particle $\alpha$. 
We normalize plane wave states according to $\langle \mathbf{p}'| \mathbf{p} \rangle = \delta(\mathbf{p}' - \mathbf{p})$. 
%Since $U_{00}(z)$ is fully symmetric under any permutation of the particles, the matrix elements ${}_{\alpha}\langle \mathbf{p}_{\alpha}, \mathbf{q}_{\alpha} | U_{0 0}(0) \lvert \mathbf{0}, \mathbf{0} \rangle$ are independent of $\alpha$.
%\ins{The Jacobi momenta $\mathbf{p}_{\alpha}$ and $\mathbf{q}_{\alpha}$ are linked to $\mathbf{p}_{\beta}$ and $\mathbf{q}_{\beta}$ with $\beta \neq \alpha$.}
% In the following we use $\mathbf{p}_{\alpha}$ and $\mathbf{q}_{\alpha}$ in the sense that $\lvert \mathbf{p}_{\alpha}, \mathbf{q}_{\alpha} \rangle_{\alpha}$ represents the same state for all $\alpha$ (i.e., $\lvert \mathbf{p}_{\alpha}, \mathbf{q}_{\alpha} \rangle_{\alpha} = \lvert \mathbf{p}_{\beta}, \mathbf{q}_{\beta} \rangle_{\beta}$ for $\alpha,\beta = 1$, 2 or 3)
Naturally, $\lvert \mathbf{p}_{\alpha}, \mathbf{q}_{\alpha} \rangle_{\alpha} = \lvert \mathbf{p}_{\beta}, \mathbf{q}_{\beta} \rangle_{\beta}$ for $\alpha,\beta = 1$, 2, or 3. The choice of $\alpha = 1$, 2, or 3 is therefore arbitrary in our definition of the elastic three-body transition amplitude ${}_{\alpha}\langle \mathbf{p}_{\alpha}, \mathbf{q}_{\alpha} | U_{0 0}(0) \lvert \mathbf{0}, \mathbf{0} \rangle$, so that we can drop the index $\alpha$ for notational compactness and write $\langle \mathbf{p}, \mathbf{q} | U_{0 0}(0) \lvert \mathbf{0}, \mathbf{0} \rangle$. This amplitude behaves as
%\\
%Therefore, the matrix elements ${}_{\alpha}\langle \mathbf{p}_{\alpha}, \mathbf{q}_{\alpha} | U_{0 0}(0) \lvert \mathbf{0}, \mathbf{0} \rangle$ are equivalent for $\alpha = 1$, 2 and 3. We can thus take $\mathbf{p} = \mathbf{p}_1$ and $\mathbf{q} = \mathbf{q}_1$ to define the elastic three-body transition amplitude $\langle \mathbf{p}, \mathbf{q} | U_{0 0}(0) \lvert \mathbf{0}, \mathbf{0} \rangle$. This amplitude behaves as
%\ins{The elastic three-body transition amplitude $\langle \mathbf{p}, \mathbf{q} | U_{0 0}(0) | \mathbf{0}, \mathbf{0} \rangle$, where $\langle \mathbf{p}, \mathbf{q} \rvert$ represents ${}_{1}\langle \mathbf{p}_1, \mathbf{q}_1 \rvert$, ${}_{2}\langle \mathbf{p}_2, \mathbf{q}_2 \rvert$ or ${}_{3}\langle \mathbf{p}_3, \mathbf{q}_3 \rvert$, is given by}
%If we define $\mathbf{p} \equiv \mathbf{p}_1$ and $\mathbf{q} \equiv \mathbf{q}_1$, the elastic three-body transition amplitude $\langle \mathbf{p}, \mathbf{q} | U_{0 0}(0) | \mathbf{0}, \mathbf{0} \rangle$ is given by
\begin{equation}\label{eq:U00_momentum}
\begin{aligned}
&\langle \mathbf{p}, \mathbf{q} | U_{0 0}(0) \lvert \mathbf{0}, \mathbf{0} \rangle = \sum_{\alpha = 1}^{3} 
\Bigg\{
{}_{\alpha}\langle \mathbf{p}_{\alpha}, \mathbf{q}_{\alpha} | T_{\alpha}(0) | \mathbf{0}, \mathbf{0}\rangle 
\\
&+ \frac{A_{\alpha}}{q_{\alpha}^2}
+ \frac{B_{\alpha}}{q_{\alpha}}
+ C_{\alpha} \, \text{ln}\bigg( \frac{q_{\alpha} \rho}{\hbar} \bigg)
+ \frac{1}{(2 \pi)^6} \mathcal{U}^{(\alpha)}(\mathbf{p}_{\alpha},\mathbf{q}_{\alpha}) \Bigg\},
\end{aligned}
\end{equation}
%\begin{equation}\label{eq:U00_momentum}
%\begin{aligned}
%&\langle \mathbf{p}, \mathbf{q} | U_{0 0}(0) \lvert \mathbf{0}, \mathbf{0} \rangle = \sum_{\substack{(\alpha, \beta, \gamma) = (1,2,3), \\ (2,3,1), (3,1,2)}} 
%\Bigg\{\delta(\mathbf{q}_{\alpha}) \langle \mathbf{p}_{\alpha} | t_{\beta \gamma}(0) | \mathbf{0}\rangle 
%\\
%&+ \frac{A_{\alpha}}{q_{\alpha}^2} + \frac{B_{\alpha}}{q_{\alpha}} + C_{\alpha} \, \text{ln}\bigg( \frac{q_{\alpha} \rho}{\hbar} \bigg) + \frac{1}{(2 \pi)^6} \mathcal{U}^{(\alpha)}(\mathbf{p}_{\alpha},\mathbf{q}_{\alpha}) \Bigg\},
%\end{aligned}
%\end{equation}
%\begin{widetext}
%\begin{eqnarray}\label{eq:U00_momentum}
%\begin{aligned}
%\langle \mathbf{p}, \mathbf{q} | U_{0 0}(0) \lvert \mathbf{0}, \mathbf{0} \rangle = &\sum_{\substack{(\alpha, \beta, \gamma) = (1,2,3), \\ (2,3,1), (3,1,2)}} 
%\Bigg\{
%\delta(\mathbf{q}_{\alpha}) \langle \mathbf{p}_{\alpha} | t_{\beta \gamma}(0) | \mathbf{0}\rangle 
%+ A_{\alpha} \frac{1}{q_{\alpha}^2}
%+ B_{\alpha} \frac{1}{q_{\alpha}}
%+ C_{\alpha} \, \text{ln}\bigg( \frac{q_{\alpha} \rho}{\hbar} \bigg)
%+ \frac{1}{(2 \pi)^6} \mathcal{U}^{(\alpha)}(\mathbf{p}_{\alpha},\mathbf{q}_{\alpha}) \Bigg\},
%\end{aligned}
%\end{eqnarray}
%\end{widetext}
where $\rho$ is an arbitrary length scale. The coefficients $A_{\alpha}$, $B_{\alpha}$, and $C_{\alpha}$ are real and depend on the masses and scattering lengths \cite{braaten2010recombinationXYZ, helfrich2010heteronuclearBBX, tan2021hypervolumeBBX, SupplMat}.
% The two-body operator $T^{(\alpha)}(z_{\mathrm{2b}})$ is defined via $T^{(\alpha)}(z_{\mathrm{2b}}) = V_{\beta \gamma} + V_{\beta \gamma} \left(z_{\mathrm{2b}}-H_0^{(\mathrm{2b})}\right)^{-1} T^{(\alpha)}(z_{\mathrm{2b}})$, where $H_0^{(\mathrm{2b})}$ is the two-body kinetic energy operator in the two-body center-of-mass frame and $z_{\mathrm{2b}}$ is the two-body energy. 
The functions $\mathcal{U}^{(\alpha)}(\mathbf{p}_{\alpha},\mathbf{q}_{\alpha})$ represent the remainder for which $\lim_{q_{\alpha} \to 0} \mathcal{U}^{(\alpha)}(\mathbf{0},\mathbf{q}_{\alpha})$ is finite \cite{noteU0finite}. So we define
\begin{equation}\label{eq:U0_def}
\mathcal{U}_0 = \sum_{\alpha = 1}^{3} \lim_{q_{\alpha} \to 0} \mathcal{U}^{(\alpha)}(\mathbf{0},\mathbf{q}_{\alpha}).
\end{equation}
This definition of $\mathcal{U}_0$ is closely related to the definition of the three-body scattering hypervolume considered in Refs.~\cite{tan2008hypervolume,mestrom2019hypervolumeSqW,mestrom2020hypervolumeVdW,tan2017hypervolume} for identical bosons and in Ref.~\cite{tan2021hypervolumeBBX} for dissimilar particles. In the Supplemental Material \cite{SupplMat} we make this connection explicit. The imaginary part of $\mathcal{U}_0$ is proportional to the three-body recombination rate due to the optical theorem for three-particle scattering \cite{schmid1974threebody, SupplMat}, whereas the real part is connected to elastic three-body scattering processes.
%\begin{equation}\label{eq:Im_U00_optical_theorem}
%\begin{aligned}
%\frac{1}{(2 \pi)^6}\mathrm{Im} &\left(\mathcal{U}_0 \right) = -\pi \sum_{\substack{(\alpha, \beta, \gamma) = (1,2,3), \\ (2,3,1), (3,1,2)}}  \sum_{d} \mu_{\alpha \beta, \gamma} q_{\gamma, d}
%\\
%&\int 
%\left\lvert{}_{\gamma}\langle \varphi_{d}^{(\alpha \beta)}, \mathbf{q}_{\gamma,d} | U_{\gamma 0}(0) | \mathbf{0}, \mathbf{0} \rangle \right\rvert^2 \,d\hat{\mathbf{q}}_{\gamma,d}.
%\end{aligned}
%\end{equation}
%Here $d$ labels the dimer states $\lvert \varphi_{d}^{(\alpha \beta)} \rangle$ consisting of particles $\alpha$ and $\beta$ whose bound state energy $E_{\mathrm{2b},d}$ fixes $q_{\gamma,d}$ via $E_{\mathrm{2b},d} = -q_{\gamma,d}^2/(2 \mu_{\alpha \beta,\gamma})$. These bound states are normalized as $\langle \varphi_{d}^{(\alpha \beta)} | \varphi_{d}^{(\alpha \beta)} \rangle = 1$.
The latter can be used to quantify the strength of an effective three-body contact interaction when modeling an ultracold quantum gas \cite{braaten1999diluteBEC, braaten2001nonuniversalBEC, braaten2002diluteBEC, braaten2006universality, tan2021hypervolumeBBX}. 

To illustrate this connection to many-body systems, we consider a dilute Bose-Bose mixture at zero temperature. A recent study \cite{tan2021hypervolumeBBX} demonstrated that the corresponding energy density $\mathcal{E}$ can be approximated by
\begin{equation}
\begin{aligned}
\mathcal{E} &= \frac{1}{6} \hbar^6 \mathcal{U}_0^{\mathrm{(BBB)}} n_{\mathrm{B}}^3 + \frac{1}{6} \hbar^6 \mathcal{U}_0^{\mathrm{(bbb)}} n_{\mathrm{b}}^3
\\
&+ \frac{1}{2} \hbar^6 \mathcal{U}_0^{\mathrm{(BBb)}} n_{\mathrm{B}}^2 n_{\mathrm{b}}
+ \frac{1}{2} \hbar^6 \mathcal{U}_0^{\mathrm{(Bbb)}} n_{\mathrm{B}} n_{\mathrm{b}}^2
\end{aligned}
\end{equation}
under the assumption that the two-body scattering lengths are negligible. Here we have denoted the two types of bosons by B and b and the corresponding number densities by $n_{\mathrm{B}}$ and $n_{\mathrm{b}}$, respectively. We have also added labels to $\mathcal{U}_0$ to distinguish those corresponding to different three-body systems. These amplitudes determine the stability of the mixture against collapse or phase separation \cite{tan2021hypervolumeBBX}. The dynamics of the mixture can be studied from the corresponding Gross-Pitaevskii equations with effective three-body contact interactions whose strengths are set by the amplitudes $\mathcal{U}_0$ \cite{tan2021hypervolumeBBX}.

\textit{$p$-wave resonance.}---To see how resonant $p$-wave interactions influence $\mathcal{U}_0$, we expand ${}_{\alpha}\langle \mathbf{p}, \mathbf{q} | T_{\alpha}(0) | \mathbf{p}', \mathbf{q}' \rangle_{\alpha}$ in the Legendre polynomials $P_l(\hat{\mathbf{p}}\cdot \hat{\mathbf{p}}')$ as
\begin{equation}
\begin{aligned}
{}_{\alpha}\langle \mathbf{p}, &\mathbf{q} | T_{\alpha}(0) | \mathbf{p}', \mathbf{q}' \rangle_{\alpha} = \langle \mathbf{q} | \mathbf{q}'\rangle 
\\
&\times \sum_{l = 0}^{\infty} (2 l + 1) P_l(\hat{\mathbf{p}}\cdot \hat{\mathbf{p}}') t_l^{(\beta \gamma)}\left(p,p',-\frac{q^2}{2 \mu_{\beta \gamma,\alpha}}\right).
\end{aligned}
\end{equation}
In contrast to identical bosons, dissimilar particles can interact via the $p$-wave component of the two-body transition amplitude, which behaves as
\begin{equation}\label{eq:T_operator_pwave}
\begin{aligned}
t_1^{(\beta \gamma)}&\left(p,p',-\frac{q^2}{2 \mu_{\beta \gamma,\alpha}}\right) = \frac{\frac{a_{1,\beta \gamma} p p'}{4 \pi^2 \mu_{\beta \gamma} \hbar^3}}{1 - \frac{1}{2} \tilde{r}_{1,\beta \gamma} a_{1,\beta \gamma} \frac{\mu_{\beta \gamma}}{\mu_{\beta \gamma,\alpha}} \frac{q^2}{\hbar^2}}
\end{aligned}
\end{equation}
%\begin{equation}\label{eq:T_operator_pwave}
%\begin{aligned}
%t_1^{(\beta \gamma)}&\left(p,p',-\frac{\hbar^2 \kappa^2}{2 \mu_{\beta \gamma}}\right) = \frac{\frac{a_{1,\beta \gamma} p p'}{4 \pi^2 \mu_{\beta \gamma} \hbar^3}}{1 - \frac{1}{2} \tilde{r}_{1,\beta \gamma} a_{1,\beta \gamma} \kappa^2}
%\end{aligned}
%\end{equation}
%\begin{equation}\label{eq:T_operator_pwave}
%\begin{aligned}
%t_1^{(\beta \gamma)}&\left(p,p',-\frac{\hbar^2 \kappa^2}{2 \mu_{\beta \gamma}}\right) 
%\\
%&= \frac{\frac{a_{1,\beta \gamma} p p'}{4 \pi^2 \mu_{\beta \gamma} \hbar^3} + O\left(p (p')^3, p' p^3\right)}{1 - \frac{1}{2} \tilde{r}_{1,\beta \gamma} a_{1,\beta \gamma} \kappa^2 + a_{1,\beta \gamma} \kappa^3 + O\left(\kappa^4\right)}
%\end{aligned}
%\end{equation}
%\begin{equation}\label{eq:T_operator_pwave}
%\begin{aligned}
%t_1^{(\mathrm{BX})}\left(p,p',-\frac{\hbar^2 \kappa^2}{2 \mu_{\mathrm{BX}}}\right) = \frac{\frac{a_{1} p p'}{4 \pi^2 \mu_{\mathrm{BX}} \hbar^3} + O\left(p (p')^3, p' p^3\right)}{1 - \frac{1}{2} \tilde{r}_{1} a_{1} \kappa^2 + a_1 \kappa^3 + O\left(\kappa^4\right)}.
%\end{aligned}
%\end{equation}
in the limit of small $p$, $p'$ and $q$ for short-range potentials \cite{taylor1972scattering}. Here $a_{1,\beta \gamma}$ is the $p$-wave scattering volume that diverges at the resonance, and $\tilde{r}_{1,\beta \gamma}>0$ is the $p$-wave effective range \cite{taylor1972scattering}. For $a_{1,\beta \gamma} \tilde{r}_{1,\beta \gamma}^3 \ll -1$, the $p$-wave state is quasibound, whereas it is bound for $a_{1,\beta \gamma} \tilde{r}_{1,\beta \gamma}^3 \gg 1$. In the latter regime, the $p$-wave dimer energy is universally described by $-\hbar^2/(\mu_{\beta \gamma} \tilde{r}_{1,\beta \gamma} a_{1,\beta \gamma})$. For van der Waals potentials, Eq.~(\ref{eq:T_operator_pwave}) is valid in the limit $|a_{1,\beta \gamma}| \to \infty$ \cite{gao2009singlechannelvdW,zhang2010pwaveFR}, which is the exact regime we concentrate on in the following.
%\insss{This result even holds for van der Waals potentials for which Eq.~(\ref{eq:T_operator_pwave}) is valid in the limit $|a_{1,\beta \gamma}| \to \infty$ as can be derived from the analysis in Ref.~\cite{zhang2010pwaveFR}.}

\begin{figure}[btp]
	\centering
	\includegraphics[width=3.4in]{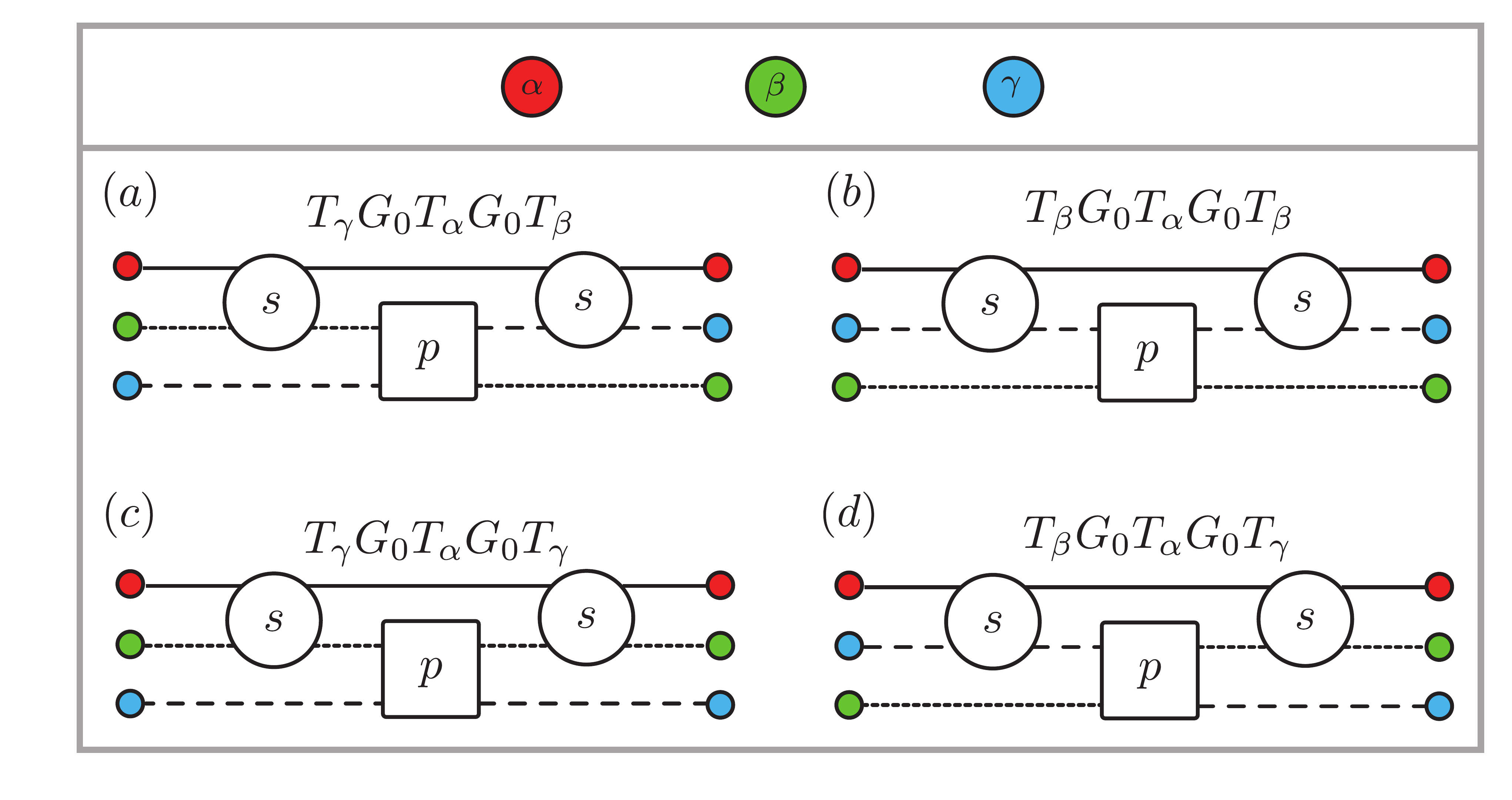}
%	\end{subfigure}
    \caption{Diagrammatic representation of the four distinct three-body scattering processes that result in the $\sqrt{-a_1}$ scaling of $\mathcal{U}_0$ due to one resonant $p$-wave interspecies interaction. In these diagrams, individual particles propagate from right to left with identities distinguished by color and line style. The vertices represent the $s$- (circles) or $p$-wave (squares) component of the two-body transition operator.
    %\ins{In these diagrams, the particles come in from the right. The particles and the corresponding propagators are distinguished by color and line style, respectively.} The vertices represent the \ins{$s$- or $p$-wave component of the two-body transition operator}.
    % The vertices represent the $T$ operators for which either the $s$-wave or $p$-wave component is relevant \ins{and we indicate this in the diagrams}.
    }
    \label{fig:diagrams_pwave_XYZ}
\end{figure}
% Figure made with Figures_diagrams.docx
% New figure made by Victor using the program Illustrator

The $p$-wave component of the two-body transition amplitude contributes to $\mathcal{U}_0$ via scattering processes containing at least three $T$ operators because the first and final $T$ operators only contribute via their $s$-wave components at zero energy. The most simple scattering events containing $p$-wave $\beta \gamma$ interactions are thus described by $\left[T_{\beta}(0) + T_{\gamma}(0)\right] G_0(0) T_{\alpha}(0) G_0(0)\left[T_{\beta}(0) + T_{\gamma}(0)\right]$. A diagrammatic representation of these scattering processes is shown in Fig.~\ref{fig:diagrams_pwave_XYZ}. In the Supplemental Material \cite{SupplMat}, we demonstrate that their contributions to $\mathcal{U}_0$ near a $p$-wave dimer resonance scale with $\sqrt{-a_{1,\beta \gamma}}$ due to an integration over the $p$-wave component in Eq.~\eqref{eq:T_operator_pwave}. For positive $a_{1,\beta \gamma}$, this integration goes over a pole, resulting in the imaginary scaling $\sqrt{-a_{1,\beta \gamma}} = i \sqrt{a_{1,\beta \gamma}}$. Terms that contain more than three $T$ operators do not contribute to the leading $\sqrt{-a_{1,\beta \gamma}}$ scaling. The dominant behavior of $\mathcal{U}_0$ close to a $p$-wave $\beta \gamma$ dimer resonance is thus given universally by
\begin{equation}\label{eq:U0_limit_large_a1_XYZ}
\begin{aligned}
\mathcal{U}_0/\sqrt{-a_{1,\mathrm{\beta \gamma}}} \underset{|a_{1,\mathrm{\beta \gamma}}| \to \infty}{=} &- 24 \sqrt{2} \pi^2 \left(\frac{a_{\mathrm{\gamma \alpha}}}{m_{\mathrm{\gamma}}} - \frac{a_{\mathrm{\alpha \beta}}}{m_{\mathrm{\beta}}}\right)^2
\\
 & \frac{\sqrt{\mu_{\mathrm{\beta \gamma,\alpha}} \mu_{\mathrm{\beta \gamma}}}}{\hbar^4 \sqrt{\tilde{r}_{1,\mathrm{\beta \gamma}}}},
\end{aligned}
\end{equation}
where the scattering lengths $a_{\mathrm{\alpha \beta}}$ and $a_{\mathrm{\gamma \alpha}}$ correspond to the $\alpha \beta$ and $\gamma \alpha$ interaction, respectively. This general result applies to three dissimilar particles with one resonant $p$-wave interaction. 

In the remainder of this Letter we focus on the BBX system, consisting of two identical bosons (B) and a distinguishable particle (X). We define $m_{\mathrm{B}}$ ($m_{\mathrm{X}}$) as the mass of particle B (X) with mass ratio $\chi \equiv m_{\mathrm{X}}/m_{\mathrm{B}}$. In the Supplemental Material \cite{SupplMat}, we derive how the coefficients $A_{\alpha}$, $B_{\alpha}$ and $C_{\alpha}$ in Eq.~\eqref{eq:U00_momentum} depend on $\chi$ and on the scattering lengths $a_{\mathrm{BB}}$ and $a_{\mathrm{BX}}$ corresponding to the BB and BX interaction, respectively.

For the BBX system with resonant $p$-wave BX interactions, the number of dominant scattering processes doubles compared with three dissimilar particles with one resonant interaction as considered in Eq.~\eqref{eq:U0_limit_large_a1_XYZ}. This results in the universal limits
%\begin{equation}\label{eq:U0_limit_large_a1neg}
%\begin{aligned}
\begin{align}\label{eq:U0_limit_large_a1neg}
\mathrm{Re}\left(\mathcal{U}_0\right)/\sqrt{|a_1|} &\underset{a_1 \to -\infty}{=} - \frac{48 \sqrt{2} \pi^2}{\sqrt{\chi (2+\chi)}} \frac{\left(\chi a_{\mathrm{BB}} - a_{\mathrm{BX}}\right)^2}{m_{\mathrm{X}}\hbar^4 \sqrt{\tilde{r}_{1}}}
\end{align}
%\end{aligned}
%\end{equation}
for $a_1<0$ and
%\begin{equation}\label{eq:U0_limit_large_a1pos}
%\begin{aligned}
\begin{align}\label{eq:U0_limit_large_a1pos}
\mathrm{Im}\left(\mathcal{U}_0\right)/\sqrt{a_1} &\underset{a_1 \to +\infty}{=} - \frac{48 \sqrt{2} \pi^2}{\sqrt{\chi (2+\chi)}} \frac{\left(\chi a_{\mathrm{BB}} - a_{\mathrm{BX}}\right)^2}{m_{\mathrm{X}}\hbar^4 \sqrt{\tilde{r}_{1}}}
\end{align}
%\end{aligned}
%\end{equation}
for $a_1>0$ \cite{SupplMat}, where we defined $a_1 \equiv a_{1,\mathrm{BX}}$ and $\tilde{r}_1 \equiv \tilde{r}_{1,\mathrm{BX}}$ for notational convenience. Clearly, the divergent behavior of $\mathcal{U}_0$ becomes stronger for smaller mass ratios $\chi$. Equation~(\ref{eq:U0_limit_large_a1pos}) can also be derived from the optical theorem, in which case one finds that the divergent behavior is caused only by three-body recombination into the weakly bound $p$-wave dimer state \cite{SupplMat}.

To study $\mathcal{U}_0$ numerically, we take a square-well potential with depth $V_0$ and range $R$ to model the BX interaction.
%\begin{equation}
%V_{\mathrm{BX}}(r)=\begin{cases}-V_0,&\mbox{$0\leq r<R$},\\0,&\mbox{$r\geq R$},\end{cases}
%\end{equation}
We fix the potential range $R$ and tune the depth $V_0$ near $2 \mu_{\mathrm{BX}} V_0 R^2/\hbar^2 = \pi^2$, which is the point where the first $p$-wave dimer state gets bound. We calculate $\mathcal{U}_0$ for this BBX system by extending the method of Ref.~\cite{mestrom2019hypervolumeSqW}, which considered three identical bosons. More specifically, starting from the AGS equations, we derive a set of integral equations for $\mathcal{U}^{(\alpha)}(\mathbf{p}_{\alpha},\mathbf{q}_{\alpha})$ \cite{SupplMat}, which we expand in spherical harmonics and Weinberg states \cite{weinberg1963expansion, mestrom2019squarewell} and discretize in $q_{\alpha}$, yielding a matrix equation that can be solved numerically. For the definition of $\mathcal{U}_0$, we fix $\rho = |a_{\mathrm{BX}}|$ in Eq.~\eqref{eq:U00_momentum}. This choice of $\rho$ is consistent with the convention of Ref.~\cite{tan2021hypervolumeBBX} when the BB interaction is set to zero as we do in our analysis presented below. %Another choice would only change real part of $\mathcal{U}_0$.
% For the definition of $\mathcal{U}_0$, we follow the convention of Refs.~\cite{braaten2002diluteBEC,braaten2006universality,tan2008hypervolume,tan2017hypervolume,mestrom2019hypervolumeSqW,mestrom2020hypervolumeVdW} for three identical bosons and fix $\rho = |a_{\mathrm{BX}}|$ in Eq.~\eqref{eq:U00_momentum}.
This convention has however no effect on the universal limits in Eqs.~\eqref{eq:U0_limit_large_a1neg} and \eqref{eq:U0_limit_large_a1pos}.

\begin{figure*}[btp]
	\begin{subfigure}
	\centering
	\includegraphics[width=3.4in]{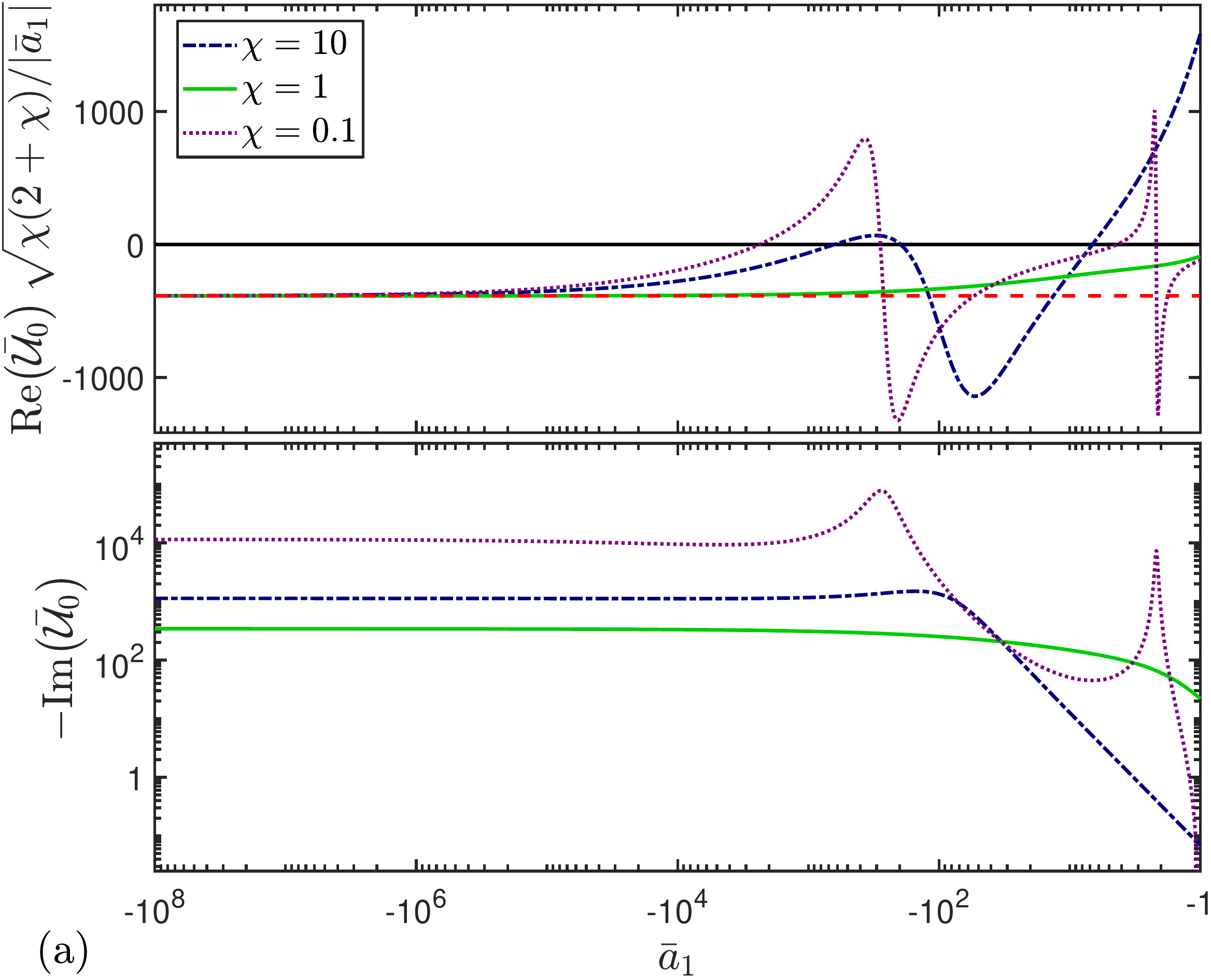}
	\end{subfigure}
\quad
	\begin{subfigure}
	\centering
	\includegraphics[width=3.4in]{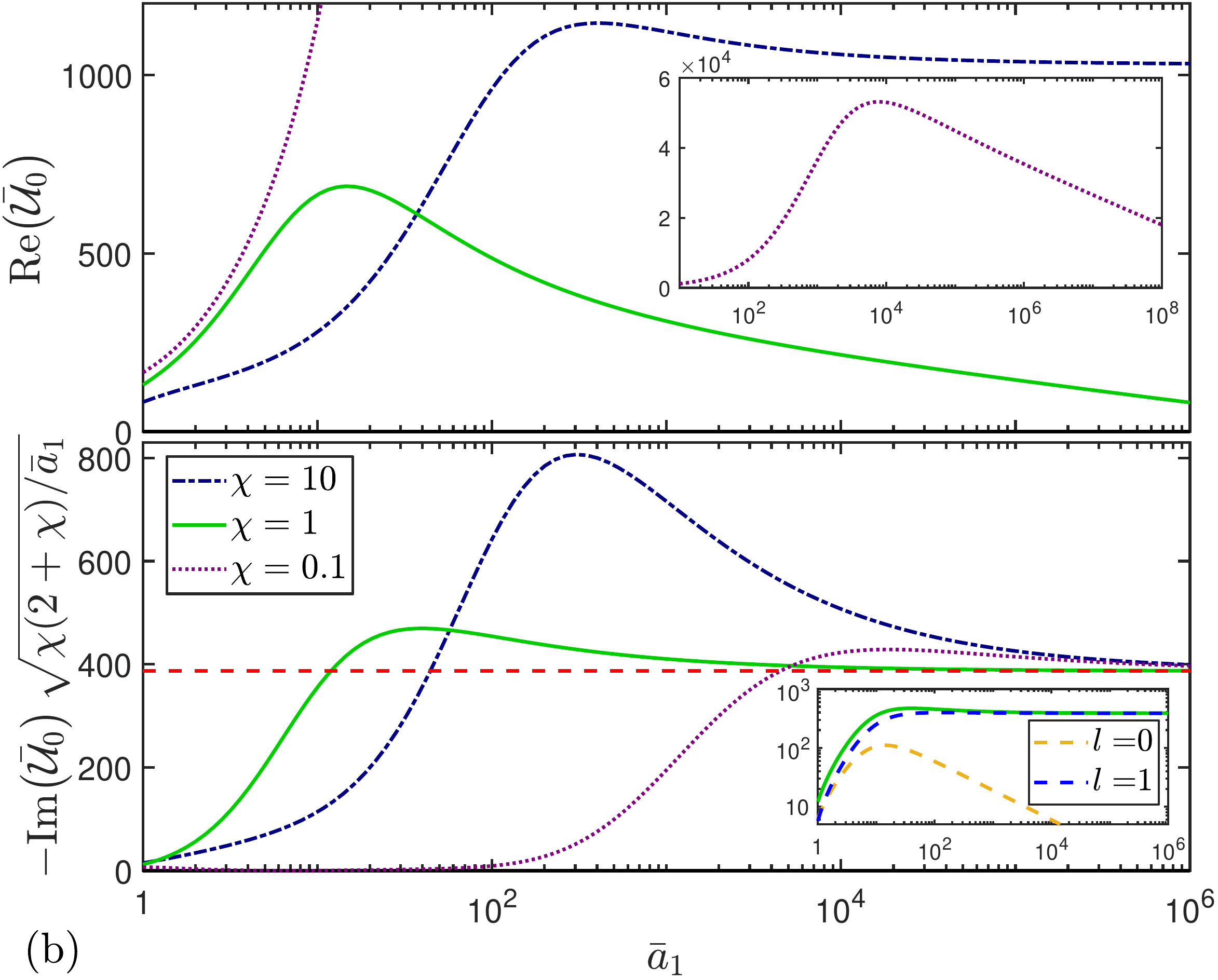}
	\end{subfigure}
    \caption{$\mathcal{U}_0$ near the first $p$-wave BX dimer resonance of the square-well potential for various mass ratios at (a) $a_1 < 0$ and (b) $a_1 > 0$. We have defined the dimensionless quantities $\bar{\mathcal{U}}_0 \equiv \mathcal{U}_0 m_{\mathrm{X}} \hbar^4/R^4$ and $\bar{a}_1 = a_1/R^3$. The BB interaction is set to zero. The red dashed lines represent Eqs.~(\ref{eq:U0_limit_large_a1neg}) and (\ref{eq:U0_limit_large_a1pos}) with $a_{\mathrm{BB}}/R = 0$, $a_{\mathrm{BX}}/R = 1$ and $\tilde{r}_1 R = 3$. The parameters $a_{\mathrm{BX}}$ and $\tilde{r}_1$ can be regarded as constants for $|\bar{a}_1|\gtrsim 10$. For $a_1>0$, $\mathrm{Im}(\mathcal{U}_0)$ is determined by the three-body recombination rate into one deep $s$-wave dimer state ($l = 0$) and one shallow $p$-wave dimer state ($l = 1$). These two contributions to $\mathrm{Im}(\mathcal{U}_0)$ are presented in the inset for $\chi = 0$. The inset for $\mathrm{Re}(\mathcal{U}_0)$ at $a_1 > 0$ demonstrates its logarithmic behavior at large $a_1$ for $\chi = 0.1$.}
    \label{fig:SqW_a1neg_a1pos_chi_0d1_1_10}
\end{figure*}

Our numerical results for $\mathcal{U}_0$ are presented in Fig.~\ref{fig:SqW_a1neg_a1pos_chi_0d1_1_10} for various mass ratios and zero BB interaction. For $a_1 \to -\infty$, $\mathrm{Re}\left(\mathcal{U}_{0}\right)$ diverges to $-\infty$ as described by Eq.~\eqref{eq:U0_limit_large_a1neg}, whereas $\mathrm{Im}\left(\mathcal{U}_{0}\right)$ stays finite. Therefore, elastic three-body scattering dominates over three-body recombination. 
%\del{Since the zero-energy limit of the three-body transition amplitude determines the ground state energy density in dilute ultracold Bose gases \cite{braaten1999diluteBEC,braaten2001nonuniversalBEC,braaten2002diluteBEC,kohler2002threebodyBEC,tan2008hypervolume}, it effectively induces a force on the gas. Therefore, the divergent behavior of $\mathrm{Re}\left(\mathcal{U}_{0}\right)$ to $-\infty$ suggests a strong effective attraction in an ultracold bosonic mixture which could destabilize the gas.}
For $a_1 \to +\infty$, Fig.~\ref{fig:SqW_a1neg_a1pos_chi_0d1_1_10} confirms the $\sqrt{a_1}$ scaling of $\mathrm{Im}\left(\mathcal{U}_{0}\right)$ as presented in Eq.~\eqref{eq:U0_limit_large_a1pos}, whereas $\mathrm{Re}\left(\mathcal{U}_{0}\right)$ diverges as $-\mathrm{ln}(a_1/R^3)$. The prefactor of this logarithmic behavior increases for smaller values of $\chi$. For large mass ratios, this behavior of $\mathrm{Re}\left(\mathcal{U}_{0}\right)$ is very subtle for the values of $a_1$ considered in Fig.~\ref{fig:SqW_a1neg_a1pos_chi_0d1_1_10}(b), since the prefactor of $-\mathrm{ln}(a_1/R^3)$ is very small. The inset in the lower panel of Fig.~\ref{fig:SqW_a1neg_a1pos_chi_0d1_1_10}(b) also demonstrates that only the part of $\mathrm{Im}\left(\mathcal{U}_{0}\right)$ that corresponds to three-body recombination into the shallow $p$-wave dimer state diverges, while all other contributions stay finite at the $p$-wave resonance.

% Figure made with code Plot_and_fit_Elastic_K3_List_BBX_v14_chi_0d1_1_10_a1neg_bar.m AND Plot_and_fit_Elastic_K3_List_BBX_v14_chi_0d1_1_10_a1pos_bar2.m

\begin{figure}[btp]
	\begin{subfigure}
	\centering
	\includegraphics[width=3.4in]{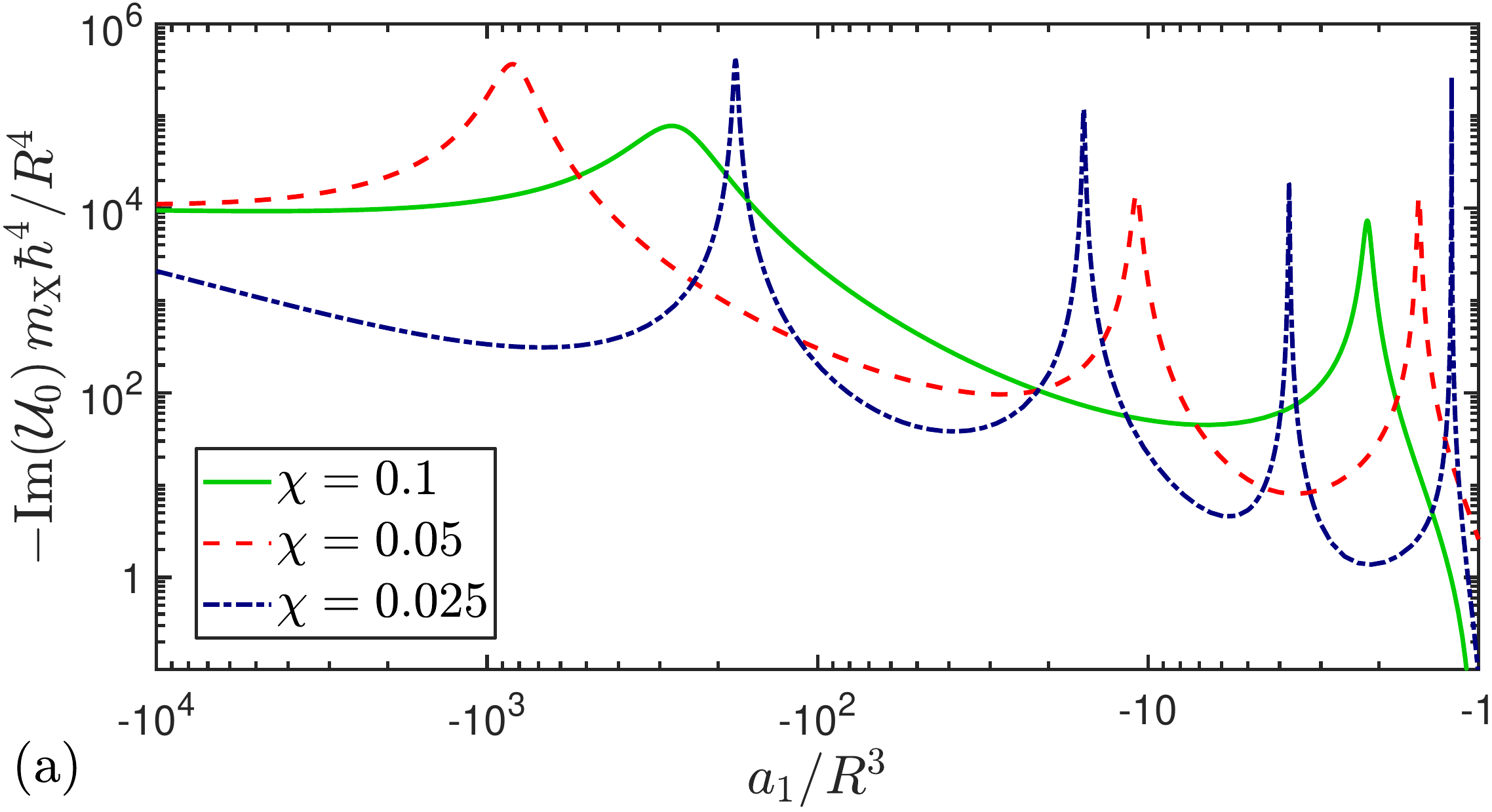}
	\end{subfigure}
\quad
	\begin{subfigure}
	\centering
	\includegraphics[width=3.4in]{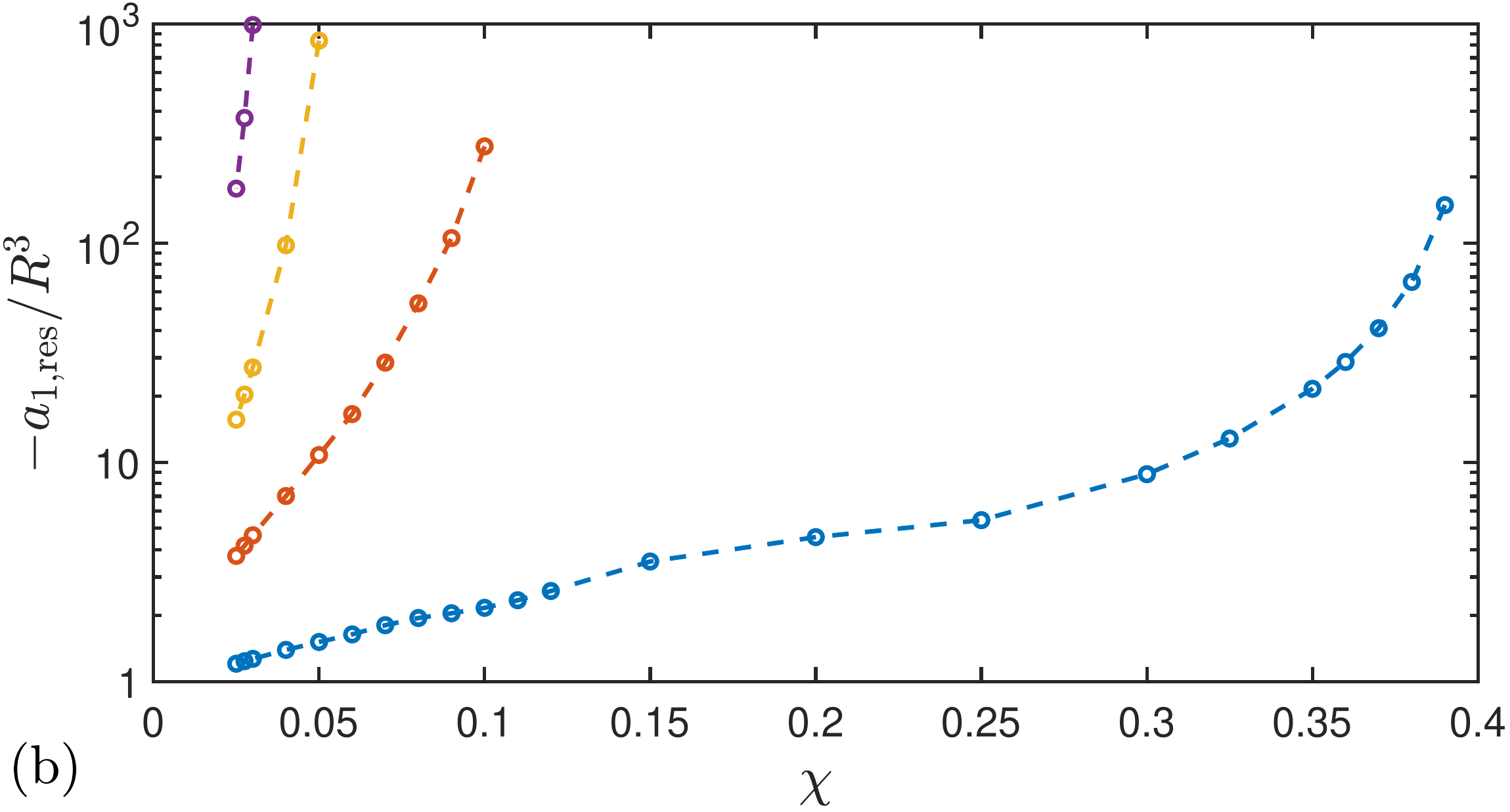}
	\end{subfigure}
    \caption{(a) $-\mathrm{Im}(\mathcal{U}_0)$ near the first $p$-wave BX dimer resonance of the square-well potential for various mass ratios $\chi \ll 1$ at $a_1<0$. The BB interaction is set to zero. (b) The $p$-wave scattering volumes $a_{1,\mathrm{res}}$ that locate the local maxima in $-\mathrm{Im}(\mathcal{U}_0)$ for $0.025\leq \chi \leq 0.4$.}
    \label{fig:SqW_a1neg_chi_0d1_0d05_0d025_TrimerRes}
\end{figure}

% At smaller values of $|a_1|$, $\mathcal{U}_{0}$ behaves nonuniversally.
When $|a_1|$ decreases, $\mathcal{U}_{0}$ starts to behave nonuniversally. For $a_1<0$, Fig.~\ref{fig:SqW_a1neg_a1pos_chi_0d1_1_10}(a) shows one trimer resonance for $\chi = 10$ near $a_1/R^3 \approx -100$ and two stronger trimer resonances for $\chi = 0.1$ near $a_1/R^3 \approx -2.17$ and $-276$ which result in clear peaks in $-\mathrm{Im}\left(\mathcal{U}_{0}\right)$.  They correspond to the three-body quasibound states with zero energy and zero total angular momentum. The trimer resonances for $\chi = 0.1$ arise from a universal long-range three-body attraction that gets stronger for smaller $\chi$ \cite{efremov2013pwaveBBX, zhu2013TwoScatteringCenters}, whereas the trimer resonance for $\chi = 10$ has not been predicted and its origin is most likely nonuniversal. 
%The trimer resonances for $\chi = 0.1$ are different from the one for $\chi = 10$, since they arise from a universal long-range three-body attraction that gets stronger for smaller $\chi$ \cite{efremov2013pwaveBBX, zhu2013TwoScatteringCenters}. 
Figure~\ref{fig:SqW_a1neg_chi_0d1_0d05_0d025_TrimerRes} demonstrates that the trimer resonances for $\chi < 1$ constitute a series whose number increases as $\chi$ decreases. This phenomenon was predicted in Ref.~\cite{efremov2013pwaveBBX} which investigated the trimer spectrum exactly on resonance. Our results show that the corresponding trimer resonances at the three-particle threshold are accompanied with large peaks in the three-body recombination rate. These resonances can even enhance this rate by a few orders of magnitude compared with the background value as shown in Fig.~\ref{fig:SqW_a1neg_chi_0d1_0d05_0d025_TrimerRes}(a). % which shift towards smaller values of $|a_1|$ as $\chi$ decreases.
%\ins{The recombination rate is a few orders of magnitude enhanced compared to its background value as can be seen from Fig.~\ref{fig:SqW_a1neg_chi_0d1_0d05_0d025_TrimerRes}(a).}
In Fig.~\ref{fig:SqW_a1neg_chi_0d1_0d05_0d025_TrimerRes}(b) we demonstrate that these trimer resonances shift towards smaller values of $|a_1|$ as $\chi$ decreases. For each trimer state there is a critical mass ratio above which the corresponding resonance has vanished. These critical mass ratios are not expected to be universal, but should depend on the details of the considered BX and BB interaction potentials.

% whose widths decrease for smaller $\chi$. NOT entirely true: comparing chi = 0.04 and 0.025 and 0.12 for example.

% For $\chi < 1$, we find a series of trimer resonances whose number increases as $\chi$ decreases. This phenomenon was predicted by Ref.~\cite{efremov2013pwaveBBX}.

%Another nonuniversal effect is the presence of local minima in $-\mathrm{Im}\left(\mathcal{U}_{0}\right)$ that arise due to destructive interference between different pathways for three-body recombination \cite{dincao2018review}. Figure~\ref{fig:SqW_a1neg_a1pos_chi_0d1_1_10}(b) shows a sharp mimimum in $-\mathrm{Im}\left(\mathcal{U}_{0}\right)$ for $\chi = 0.1$ at $a_1/R^3 = 6.9$. We suspect that it arises from such interference effects.

% Figure made with code Plot_and_fit_Elastic_K3_List_BBX_v14_chi_0d1_0d05_0d025_ImU0.m AND Plot_BBX_trimer_resonances_positions_SqW.m
% You can analyze the local maxima with my code Plot_and_fit_Elastic_K3_List_BBX_v15_Analyze_Efremov_3b.m

%\ins{The universal limit presented in Eqs.~\eqref{eq:U0_limit_large_a1neg} and \eqref{eq:U0_limit_large_a1pos} for the BBX system can be generalized to a system consisting of three dissimilar particles X, Y and Z, since it arises via scattering events that include only one resonant $p$-wave interaction.} For this XYZ system with a resonant $p$-wave YZ interaction, the dominant scattering events are described by $\left(T_{\mathrm{Y}}(0) + T_{\mathrm{Z}}(0)\right) G_0(0) T_{\mathrm{X}}(0) G_0(0)\left(T_{\mathrm{Y}}(0) + T_{\mathrm{Z}}(0)\right)$. This results in the universal formula
%\begin{equation}\label{eq:U0_limit_large_a1_XYZ}
%\begin{aligned}
%\mathcal{U}_0/\sqrt{-a_{1,\mathrm{YZ}}} \underset{|a_{1,\mathrm{YZ}}| \to \infty}{=} &- 24 \sqrt{2} \pi^2 \left(\frac{a_{\mathrm{ZX}}}{m_{\mathrm{Z}}} - \frac{a_{\mathrm{XY}}}{m_{\mathrm{Y}}}\right)^2
%\\
% & \frac{\sqrt{\mu_{\mathrm{YZ,X}} \mu_{\mathrm{YZ}}}}{\hbar^4 \sqrt{\tilde{r}_{1,\mathrm{YZ}}}},
%\end{aligned}
%\end{equation}
%where $a_{1,\mathrm{YZ}}$ and $\tilde{r}_{1,\mathrm{YZ}}$ are the $p$-wave scattering volume and effective range corresponding to the YZ interaction, respectively. The scattering lengths $a_{\mathrm{XY}}$ and $a_{\mathrm{ZX}}$ correspond to the XY and ZX interaction, respectively.

\textit{Comparison with $s$-wave resonances.}---The universal behavior of $\mathcal{U}_0$ for the BBX system near a $p$-wave dimer resonance differs from the behavior near an $s$-wave dimer resonance (i.e., $|a_{\mathrm{BX}}| \to \infty$) where the Efimov effect \cite{efimov1970energy, efimov1971weakly, efimov1973energy, braaten2006universality, naidon2017review, greene2017review, dincao2018review, helfrich2010heteronuclearBBX, mikkelsen2015K3BBX} causes $\mathcal{U}_{0}$ to be a log-periodic function of $a_{\mathrm{BX}}$ attached to an $a_{\mathrm{BX}}^4$ scaling.
%For strong $s$-wave interactions ($|a_{\mathrm{BX}}| \to \infty$), $\mathcal{U}_{0}$ scales as $a_{\mathrm{BX}}^4$ \cite{SupplMat}. 
The latter is nonperturbative, while the $\sqrt{-a_{1,\mathrm{BX}}}$ scaling for resonant $p$-wave interactions only involves three-body collisions described by three $T$ operators. In addition, three-body recombination into deeply bound dimer states also contributes to the leading $a_{\mathrm{BX}}^4$ scaling on both sides of an $s$-wave dimer resonance, whereas such contributions are nondivergent for resonant $p$-wave interactions. For completeness, we present an overview of the universal behavior of $\mathcal{U}_{0}$ near an $s$-wave BX dimer resonance, of which most is already known, in the Supplemental Material \cite{SupplMat}.

\textit{Outlook.}---In the $p$-wave universal regime [Eqs.~(\ref{eq:U0_limit_large_a1neg}) and (\ref{eq:U0_limit_large_a1pos})], $\mathcal{U}_0$ of the BBX system diverges at a point where the $s$-wave scattering lengths are generally finite.
This implies that three-body scattering dominates over two-body scattering at zero energy in an ultracold $p$-wave resonant mixture and could therefore strongly alter previous predictions for the phase diagram \cite{radzihovsky2009pwaveResonantBoseMixture, radzihovsky2011pwaveResonantBoseMixture, li2019pwaveResonantBoseMixture}.
% and should thus be included when determining the phase diagram.
%for large enough values of $|a_1|$ in an ultracold mixture of B and X particles \inss{and should thus be included when determining the phase diagram for ultracold $p$-wave resonant mixtures}. 
%in an ultracold mixture of B and X particles. The corresponding densities and scattering lengths determine how large $|a_1|$ needs to be to achieve this dominant three-body scattering. In addition, two-body $p$-wave collisions are suppressed if the temperature is sufficiently low. 
%The stability of ultracold Bose-Bose and Bose-Fermi mixtures near a $p$-wave interspecies dimer resonance could thus be strongly influenced by three-body collisions. Since the zero-energy limit of the three-body transition amplitude determines the ground state energy density in dilute ultracold Bose gases
%\cite{braaten1999diluteBEC,braaten2001nonuniversalBEC,braaten2002diluteBEC,kohler2002threebodyBEC,tan2008hypervolume}, it effectively induces a force on the gas. Therefore, the divergent behavior of $\mathrm{Re}\left(\mathcal{U}_{0}\right)$ to $-\infty$ as $a_1 \to -\infty$ suggests a strong effective attraction in an ultracold mixture which could destabilize the gas \ins{and strongly alter the phase diagram.}
In particular, the divergent behavior of $\mathrm{Re}\left(\mathcal{U}_{0}\right)$ to $-\infty$ as $a_1 \to -\infty$ suggests a strong effective attraction in ultracold mixtures which could have a destabilizing effect.
% and strongly alter the phase diagram.

%%% Old paragraph: \textit{Collisions in ultracold gases.}---Equations~(\ref{eq:U0_limit_large_a1neg}) and (\ref{eq:U0_limit_large_a1pos}) show that $\mathcal{U}_0$ of the BBX system diverges at a point where the $s$-wave scattering lengths are generally finite. This implies that three-body scattering dominates over two-body scattering at zero energy for large enough values of $|a_1|$ in an ultracold mixture of B and X particles. The corresponding densities and scattering lengths determine how large $|a_1|$ needs to be to achieve this dominant three-body scattering. In addition, two-body $p$-wave collisions are suppressed if the temperature is sufficiently low. The stability of ultracold Bose-Bose and Bose-Fermi mixtures near a $p$-wave interspecies dimer resonance could thus be strongly influenced by three-body collisions. Since the zero-energy limit of the three-body transition amplitude determines the ground state energy density in dilute ultracold Bose gases\cite{braaten1999diluteBEC,braaten2001nonuniversalBEC,braaten2002diluteBEC,kohler2002threebodyBEC,tan2008hypervolume}, it effectively induces a force on the gas. Therefore, the divergent behavior of $\mathrm{Re}\left(\mathcal{U}_{0}\right)$ to $-\infty$ as $a_1 \to -\infty$ suggests a strong effective attraction in an ultracold mixture which could destabilize the gas \ins{and strongly alter the phase diagram.}

%\inss{Furthermore, Eq.~(\ref{eq:U0_limit_large_a1neg}) offers a novel approach to realize quantum gases dominated by elastic three-body scattering.} 
We note that divergent behavior of $\mathrm{Re}\left(\mathcal{U}_{0}\right)$ can also occur when the potentials support a three-body bound state at zero energy whose total angular momentum is zero. 
%\insss{, but support no dimer states to which three particles can recombine.}
% In this case, $\mathrm{Re}\left(\mathcal{U}_{0}\right)$ only diverges in the absence of dimer states to which three particles can recombine.
This is only possible in the absence of dimer states to which three particles can recombine. However, our universal result in Eq.~(\ref{eq:U0_limit_large_a1neg}) applies even when deeply bound dimer states exist. This remarkable property makes the $p$-wave dimer resonance a promising tool to realize a divergent $\mathrm{Re}\left(\mathcal{U}_{0}\right)$ in atomic systems that typically support many dimer states.

On the other hand, the imaginary part of $\mathcal{U}_0$ is experimentally observable in a trapped ultracold atomic gas by measuring the atom loss from the trap as a function of time. We identify the following conditions that are required to observe the universal behavior of $\mathrm{Im}\left(\mathcal{U}_0\right)$ in Eq.~\eqref{eq:U0_limit_large_a1pos}. First, the interspecies Feshbach resonance needs to be broad enough to accurately tune $a_1$ up to large values. Such a broad $p$-wave Feshbach resonance was found in a Bose-Bose mixture of ${}^{85}$Rb and ${}^{87}$Rb atoms \cite{dong2016pwaveFRRb85Rb87}. Secondly, the gas needs to be cold enough to neglect temperature effects. 
%\del{Therefore, the thermal energy $k_B T$ needs to be much smaller than the $p$-wave dimer binding energy $\hbar^2/(\mu_{\mathrm{BX}} \tilde{r}_1 a_1)$.}
We expect such temperature effects to be strong due to another dominant contribution to the three-body recombination rate at positive three-body energies $E$ and $a_1>0$, scaling as $E^2 a_1^{5/2} /\sqrt{\tilde{r}_1}$, which is similar for three identical fermions \cite{jonalasinio2008pwaveFRsystemFFF}. Therefore, the thermal energy needs to be much smaller than $\hbar^2|\chi a_{\mathrm{BB}} - a_{\mathrm{BX}}|/( m_{\mathrm{X}} a_1)$ to observe the behavior in Eq.~\eqref{eq:U0_limit_large_a1pos}. Specifically for the broad $p$-wave Feshbach resonance in a ${}^{85}\mathrm{Rb}$--${}^{87}\mathrm{Rb}$ mixture at a magnetic field of 823.3 G \cite{dong2016pwaveFRRb85Rb87}, we find that $\hbar^2|\chi a_{\mathrm{BB}} - a_{\mathrm{BX}}|/(k_B m_{\mathrm{X}} a_1) \approx 200$~nK for $\mathrm{BBX} = {}^{85}\mathrm{Rb} {}^{85}\mathrm{Rb} {}^{87}\mathrm{Rb}$ and $20$ nK for $\mathrm{BBX} = {}^{87}\mathrm{Rb} {}^{87}\mathrm{Rb} {}^{85}\mathrm{Rb}$ \cite{noteRb8587ScatteringLengths}, where we take $a_1/r_{\mathrm{vdW}}^3 = 10^4$ according to Fig.~\ref{fig:SqW_a1neg_a1pos_chi_0d1_1_10}(b) and $r_{\mathrm{vdW}}$ is the van der Waals length scale characterizing the range of the interatomic BX interaction \cite{chin2010feshbach}. Since the magnitude of $\mathrm{Im}\left(\mathcal{U}_0\right)$ in Eq.~\eqref{eq:U0_limit_large_a1pos} for $\mathrm{BBX} = {}^{85}\mathrm{Rb} {}^{85}\mathrm{Rb} {}^{87}\mathrm{Rb}$ is more than 100 times larger than the one for $\mathrm{BBX} = {}^{87}\mathrm{Rb} {}^{87}\mathrm{Rb} {}^{85}\mathrm{Rb}$ \cite{noteRb8587ScatteringLengths}, the total decay rate is primarily determined by the ${}^{85}\mathrm{Rb} {}^{85}\mathrm{Rb} {}^{87}\mathrm{Rb}$ system close to the $p$-wave dimer resonance. Therefore, it suffices to consider temperatures that are well below $200$ nK to neglect temperature effects on the total decay rate when tuning $a_1/r_{\mathrm{vdW}}^3$ up to $10^4$.
In addition, the behavior in Eq.~\eqref{eq:U0_limit_large_a1pos} dominates over other contributions to the total recombination rate at zero energy when $a_1$ is chosen large enough. Estimating these contributions generally requires accurate interaction models that
account for the exact three-atom spin structure.
Furthermore, it is beneficial to take $\chi \simeq 1$, since Fig.~\ref{fig:SqW_a1neg_a1pos_chi_0d1_1_10}(b) demonstrates that the universal limit of $\mathrm{Im}\left(\mathcal{U}_0\right)$ is approached faster for $\chi = 1$ than for $\chi\ll 1$ or $\chi \gg 1$. Fortunately, a good candidate is readily available in a mixture of ${}^{85}$Rb and ${}^{87}$Rb. Lastly, three-body recombination into the shallow $p$-wave dimer state only gives rise to atom loss when the depth of the trapping potential is smaller than the binding energy of this dimer state. Tuning $a_1$ to large values thus provides an efficient way to create weakly bound $p$-wave molecules that remain trapped, since only the three-body recombination rate into the shallow dimer state diverges on resonance.
% See Matlab code Estimates_Temperature_K3_atomic_species_pwave.m

Finally, we note that the magnetic dipole-dipole interaction between the valence electrons of alkali-metal atoms splits a $p$-wave Feshbach resonance into two \cite{ticknor2004pwaveFR,ahmedbraun2021pwaveFR}. This splitting depends on the quantum number corresponding to the projection of the molecular orbital angular momentum onto the magnetic field axis. Therefore, the universal limits in Eqs.~\eqref{eq:U0_limit_large_a1_XYZ}--\eqref{eq:U0_limit_large_a1pos} will have an additional dependence on this quantum number for these atoms. Nevertheless, we expect that the $\sqrt{-a_1}$ scaling of $\mathcal{U}_0$ is unchanged in the regime where the $p$-wave dimer binding energy is well described by $\hbar^2/(\mu_{\mathrm{BX}} \tilde{r}_1 a_1)$. 
%\del{(Or do we expect no dependenc on $m_l$, since it changes the result only at very weak binding energies? Jin 2007 is super useful, Fuchs 2008 maybe also)}
%\del{(Should I derive the result including the magnetic dipole-dipole interaction? Note that it is not spherically symmetric. Can you define a p-wave scattering volume?)}

\textit{Conclusion.}---We have studied zero-energy scattering for mixed three-body systems with resonant $p$-wave interspecies interactions. We have found a universal relation between the three-body transition amplitude $\mathcal{U}_0$ and $p$-wave scattering volume $a_1$, behaving as $\mathcal{U}_0 \propto \sqrt{-a_1}$.
%We have shown that the three-body transition amplitude $\mathcal{U}_0$ scales universally as $\sqrt{-a_1}$.
For $a_1>0$, $\mathcal{U}_0$ is dominated by three-body recombination into the weakly bound $p$-wave dimer state. For $a_1<0$, the dominant contribution comes from elastic three-body scattering processes that involve three successive two-body collisions. The limit $a_1 \to -\infty$ thus offers a special regime in which elastic three-body scattering dominates over two-body scattering and three-body recombination in ultracold mixtures. This general effect could significantly impact the phase diagram of these gases. For smaller values of $|a_1|$, $\mathcal{U}_0$ of the BBX system is influenced by a series of trimer states consisting of one light particle (X) and two heavy bosons (B). This could be relevant for nuclear systems for which other trimer states bound by strong $p$-wave interactions have been found \cite{ji2016halonuclei3body,hammer2017halonucleiEFT}.

\textit{Acknowledgments.}---We thank Denise Ahmed-Braun, Gijs Groeneveld, and Silvia Musolino for discussions. This research is financially supported by the Netherlands Organisation for Scientific Research (NWO) under Grant No. 680-47-623. V.E.C. acknowledges additional financial support from Provincia Autonoma di Trento and the Italian MIUR under the PRIN2017 projectCEnTraL.

\bibliographystyle{apsrev}
\bibliography{Bibliography}

\begin{thebibliography}{67}
\bibitem[{not({\natexlab{a}})}]{noteU0finite}
\bibinfo{note}{Generally, $\mathcal{U}_0$ is finite. Exceptions include
  resonant $s$- and $p$-wave interactions and potentials that support a
  three-body bound state at zero energy whose total angular momentum is zero.}
\end{thebibliography}

%%%%%%%%%%%%%%%% SUPPLEMENTARY

%\pagebreak
%\newpage
\clearpage

\onecolumngrid
\begin{center}
  \textbf{\large Supplemental Material: ``Three-body universality in ultracold $p$-wave resonant mixtures''}\\[.2cm]
  P. M. A. Mestrom,$^{1}$ V. E. Colussi,$^{1,2}$ T. Secker,$^{1}$ J.-L. Li,$^{1}$ and S. J. J. M. F. Kokkelmans$^1$\\[.1cm]
  {\itshape ${}^1$Eindhoven University of Technology, P.~O.~Box 513, 5600 MB Eindhoven, The Netherlands\\}
  {\itshape ${}^2$INO-CNR BEC Center and Dipartimento di Fisica, Universit\`{a} di Trento, 38123 Povo, Italy\\}
%(Dated: \today)\\[1cm]
\end{center}
\twocolumngrid

\setcounter{equation}{0}
\setcounter{figure}{0}
\setcounter{page}{1}
\renewcommand{\theequation}{S\arabic{equation}}
\renewcommand{\thefigure}{S\arabic{figure}}
\renewcommand{\thetable}{S\arabic{table}}  
\renewcommand{\bibnumfmt}[1]{[S#1]}

\section{Connection to the three-body scattering hypervolume}
\label{sec:U0_vs_hypervolume_mixture}

A recent study \cite{tan2021hypervolumeBBX} defined the three-body scattering hypervolume $D$ by
\begin{equation}
D \equiv 3 \sqrt{3} \frac{\sqrt{m_{1} + m_{2} + m_{3}} \left(m_{1} m_{2} m_{3}\right)^{3/2}}{\left(m_{1} m_{2} + m_{2} m_{3} + m_{3} m_{1}\right)^2} \, \tilde{D},
\end{equation}
where $\tilde{D}$ connects to $\mathcal{U}_0$ via
\begin{equation}\label{eq:tildeD_vs_U0}
\begin{aligned}
\tilde{D}/\hbar^4 &= \mathcal{U}_0 + \sum_{\substack{(\alpha, \beta, \gamma) = (1,2,3), \\ (2,3,1),\,(3,1,2)}} (2 \pi)^6 \Bigg\{ C_{\alpha} \, \mathrm{ln}\left(\frac{\rho}{|a_{\beta \gamma}|}\right) 
\\
&\phantom{=} + \frac{\mu_{\beta \gamma}}{\mu_{\beta \gamma,\alpha}} \frac{a_{\alpha \beta} + a_{\gamma \alpha}}{4 \pi^2 \hbar} \frac{\partial^2 t_0^{(\beta \gamma)}(p,0,0)}{\partial p^2}\Bigg|_{p=0} \Bigg\}.
\end{aligned}
\end{equation}
%Here we defined the two-body transition operator $t_{\beta \gamma}(z_{\mathrm{2b}})$ via $t_{\beta \gamma}(z_{\mathrm{2b}}) = V_{\beta \gamma} + V_{\beta \gamma} \left(z_{\mathrm{2b}}-H_0^{(\mathrm{2b})}\right)^{-1} t_{\beta \gamma}(z_{\mathrm{2b}})$ \cite{taylor1972scattering}, where $z_{\mathrm{2b}}$ is the two-body energy and $H_0^{(\mathrm{2b})}$ is the two-body kinetic energy operator in the two-body center-of-mass frame. 
%where $a_{\alpha} \equiv a_{\beta \gamma}$ with $\left\{\alpha,\beta,\gamma\right\} = \left\{1,2,3\right\}$, $\left\{2,3,1\right\}$ or $\left\{3,1,2\right\}$.
Near an interspecies $p$-wave dimer resonance $\tilde{D}$ diverges in the same way as $\hbar^4 \mathcal{U}_0$ because the other terms in Eq.~\eqref{eq:tildeD_vs_U0} do not depend on the $p$-wave component of the pairwise interactions.

\section{Three-body scattering for resonant $p$-wave interactions}

In this section we analyze how the terms $\left[T_{\beta}(0) + T_{\gamma}(0)\right] G_0(0) T_{\alpha}(0) G_0(0)\left[T_{\beta}(0) + T_{\gamma}(0)\right]$ contribute to $\mathcal{U}_0$ for $|a_{1,\beta \gamma}| \to \infty$. %\del{For this purpose, we first connect the three-body operator $T_{\alpha}(z)$ to the usual two-body transition operator $t_{\beta \gamma}(z_{\mathrm{2b}})$ that is defined via $t_{\beta \gamma}(z_{\mathrm{2b}}) = V_{\beta \gamma} + V_{\beta \gamma} \left(z_{\mathrm{2b}}-H_0^{(\mathrm{2b})}\right)^{-1} t_{\beta \gamma}(z_{\mathrm{2b}})$ \cite{taylor1972scattering}. Here $z_{\mathrm{2b}}$ is the two-body energy and $H_0^{(\mathrm{2b})}$ is the two-body kinetic energy operator in the two-body center-of-mass frame.} Consequently,
We note that
\begin{equation}
\begin{aligned}
{}_{\alpha}\langle \mathbf{p}, \mathbf{q} &| T_{\alpha}(0) | \mathbf{p}', \mathbf{q}' \rangle_{\alpha} = \langle \mathbf{q} | \mathbf{q}'\rangle \langle \mathbf{p} | t_{\beta \gamma}\left(-\frac{q^2}{2 \mu_{\beta \gamma,\alpha}}\right) | \mathbf{p}'\rangle,
\end{aligned}
\end{equation}
where we connected the three-body operator $T_{\alpha}(z)$ to the usual two-body transition operator $t_{\beta \gamma}(z_{\mathrm{2b}})$ that is defined via $t_{\beta \gamma}(z_{\mathrm{2b}}) = V_{\beta \gamma} + V_{\beta \gamma} G_0^{(\mathrm{2b})}(z_{\mathrm{2b}}) t_{\beta \gamma}(z_{\mathrm{2b}})$ \cite{taylor1972scattering}. Here $z_{\mathrm{2b}}$ is the two-body energy, $G_0^{(\mathrm{2b})}(z_{\mathrm{2b}}) = \left(z_{\mathrm{2b}}-H_0^{(\mathrm{2b})}\right)^{-1}$ and $H_0^{(\mathrm{2b})}$ is the two-body kinetic energy operator in the two-body center-of-mass frame.
We consider spherically symmetric potentials for which
\begin{equation}
\langle \mathbf{p} | t_{\beta \gamma}(z_{\mathrm{2b}}) | \mathbf{p}'\rangle = \sum_{l = 0}^{\infty} (2 l + 1) P_l(\hat{\mathbf{p}}\cdot \hat{\mathbf{p}}') t_l^{(\beta \gamma)}\left(p,p',z_{\mathrm{2b}}\right),
\end{equation}
\begin{equation}
t_{l}^{(\beta \gamma)}\left(p,p',z_{\mathrm{2b}}\right) = t_{l}^{(\beta \gamma)}\left(p',p,z_{\mathrm{2b}}\right),
\end{equation}

\begin{equation}\label{eq:tlneq0_zero_p}
t_{l\neq 0}^{(\beta \gamma)}\left(0,p',z_{\mathrm{2b}}\right) = 0,
\end{equation}
\begin{equation}\label{eq:t0_small_p_pa_z}
\begin{aligned}
 t_0^{(\beta \gamma)}&\left(p,p',-\frac{\hbar^2 \kappa^2}{2 \mu_{\beta \gamma}}\right) = \frac{\frac{a_{\beta \gamma}}{4 \pi^2 \mu_{\beta \gamma} \hbar} + O\left(p^2, (p')^2\right)}{1 - a_{\beta \gamma} \kappa + O\left(\kappa^2\right)}
\end{aligned}
\end{equation}
and
\begin{equation}\label{eq:t1_small_p_pa_z}
\begin{aligned}
 t_1^{(\beta \gamma)}&\left(p,p',-\frac{\hbar^2 \kappa^2}{2 \mu_{\beta \gamma}}\right) = \left(\frac{a_{1,\beta \gamma} p p'}{4 \pi^2 \mu_{\beta \gamma} \hbar^3} + O\left(p (p')^3, p' p^3\right) \right)
\\
&\times
\frac{1}{1 - \frac{1}{2} \tilde{r}_{1,\beta \gamma} a_{1,\beta \gamma} \kappa^2 + a_{1,\beta \gamma} \kappa^3 + O\left(\kappa^4\right)}.
\end{aligned}
\end{equation}
As a consequence, we find for arbitrary constants $x$ and $y$ that
\begin{equation}\label{eq:int_1/q^2_t1_dq_a1abs_to_inf}
\begin{aligned}
\Bigg(\int_0^{Q} &\frac{1}{q^2}  t_1^{(\beta \gamma)}\left(x q,y q,-\frac{q^2}{2 \mu_{\beta \gamma,\alpha}} + i 0\right) \,dq \Bigg)\frac{1}{\sqrt{-a_{1,\beta \gamma}}}
\\
&\underset{a_{1,\beta \gamma} \to \pm\infty}{=} - \frac{\sqrt{2}}{8 \pi} \, x y \,  \sqrt{\frac{\mu_{\beta \gamma,\alpha}}{\mu_{\beta \gamma}}}\frac{1}{\mu_{\beta \gamma} \hbar^2 \sqrt{\tilde{r}_{1,\beta \gamma}}},
\end{aligned}
\end{equation}
where $Q$ is a positive upper limit that can be chosen to be arbitrarily small.

Next, we analyze the four matrix elements
\newpage
\begin{widetext}
\begin{align}
\begin{split}\label{eq:TbG0TaG0Tb_momentum_space}
{}_{\beta} \langle \mathbf{p}, \mathbf{q} | &T_{\beta}(0) G_0(0) T_{\alpha}(0) G_0(0) T_{\beta}(0) | \mathbf{0}, \mathbf{0} \rangle
= 4 \mu_{\gamma \alpha} \int \frac{1}{q'^2}
\frac{1}{\frac{1}{\mu_{\gamma \alpha}} q'^2 + \frac{2}{m_{\gamma}} \mathbf{q}\cdot\mathbf{q}' + \frac{1}{\mu_{\beta \gamma}} q^2}
\\
&\langle \mathbf{p} | t_{\gamma \alpha}\left(-\frac{q^2}{2 \mu_{\gamma \alpha, \beta}}\right) | -\mathbf{q}' - \frac{\mu_{\gamma \alpha}}{m_{\gamma}} \mathbf{q} \rangle 
\langle  \mathbf{q} + \frac{\mu_{\beta \gamma}}{m_{\gamma}} \mathbf{q}' | t_{\beta \gamma}\left(-\frac{q'^2}{2 \mu_{\beta \gamma, \alpha}}\right) |\frac{\mu_{\beta \gamma}}{m_{\gamma}} \mathbf{q}' \rangle
\langle -\mathbf{q}' | t_{\gamma \alpha}(0) | \mathbf{0} \rangle \,d\mathbf{q}',
\end{split}
%%%%%%%%%%%%%%%%%%%%%%%%%%%%%%%%%%%%%%%%%%%%%%%%%%%
\\
\begin{split}\label{eq:TgG0TaG0Tg_momentum_space}
{}_{\gamma} \langle \mathbf{p}, \mathbf{q} | &T_{\gamma}(0) G_0(0) T_{\alpha}(0) G_0(0) T_{\gamma}(0) | \mathbf{0}, \mathbf{0} \rangle
= 4 \mu_{\alpha \beta} \int \frac{1}{q'^2} \frac{1}{\frac{1}{\mu_{\alpha \beta}} q'^2 + \frac{2}{m_{\beta}} \mathbf{q}\cdot\mathbf{q}' + \frac{1}{\mu_{\beta \gamma}} q^2}
\\
&\langle \mathbf{p} | t_{\alpha \beta}\left(-\frac{q^2}{2 \mu_{\alpha \beta, \gamma}}\right) | \mathbf{q}' + \frac{\mu_{\alpha \beta}}{m_{\beta}} \mathbf{q} \rangle
\langle  -\mathbf{q} - \frac{\mu_{\beta \gamma}}{m_{\beta}} \mathbf{q}' | t_{\beta \gamma}\left(-\frac{q'^2}{2 \mu_{\beta \gamma, \alpha}}\right) | - \frac{\mu_{\beta \gamma}}{m_{\beta}}\mathbf{q}' \rangle
\langle \mathbf{q}' | t_{\alpha \beta}(0) | \mathbf{0} \rangle
 \,d\mathbf{q}',
 \end{split}
 %%%%%%%%%%%%%%%%%%%%%%%%%%%%%%%%%%%%%%%%%%%%%%%%%%%
 \\
 \begin{split}\label{eq:TbG0TaG0Tg_momentum_space}
 {}_{\beta} \langle \mathbf{p}, \mathbf{q} | &T_{\beta}(0) G_0(0) T_{\alpha}(0) G_0(0) T_{\gamma}(0) | \mathbf{0}, \mathbf{0} \rangle 
 = 4 \mu_{\alpha \beta} \int \frac{1}{q'^2} \frac{1}{\frac{1}{\mu_{\gamma \alpha}} q'^2 + \frac{2}{m_{\gamma}} \mathbf{q}\cdot\mathbf{q}' + \frac{1}{\mu_{\beta \gamma}} q^2}
 \\
& \langle \mathbf{p} | t_{\gamma \alpha}\left(-\frac{q^2}{2 \mu_{\gamma \alpha, \beta}}\right) | -\mathbf{q}' - \frac{\mu_{\gamma \alpha}}{m_{\gamma}} \mathbf{q} \rangle 
\langle  \mathbf{q} + \frac{\mu_{\beta \gamma}}{m_{\gamma}} \mathbf{q}' | t_{\beta \gamma}\left(-\frac{q'^2}{2 \mu_{\beta \gamma, \alpha}}\right) |-\frac{\mu_{\beta \gamma}}{m_{\beta}} \mathbf{q}' \rangle
\langle \mathbf{q}' | t_{\alpha \beta}(0) | \mathbf{0} \rangle  \,d\mathbf{q}',
\end{split}
 %%%%%%%%%%%%%%%%%%%%%%%%%%%%%%%%%%%%%%%%%%%%%%%%%%%
 \\
 \begin{split}\label{eq:TgG0TaG0Tb_momentum_space}
 {}_{\gamma} \langle \mathbf{p}, \mathbf{q} | &T_{\gamma}(0) G_0(0) T_{\alpha}(0) G_0(0) T_{\beta}(0) | \mathbf{0}, \mathbf{0} \rangle
 = 4 \mu_{\gamma \alpha} \int \frac{1}{q'^2} \frac{1}{\frac{1}{\mu_{\alpha \beta}} q'^2 + \frac{2}{m_{\beta}} \mathbf{q}\cdot\mathbf{q}' + \frac{1}{\mu_{\beta \gamma}} q^2}
 \\
&\langle \mathbf{p} | t_{\alpha \beta}\left(-\frac{q^2}{2 \mu_{\alpha \beta, \gamma}}\right) | \mathbf{q}' + \frac{\mu_{\alpha \beta}}{m_{\beta}} \mathbf{q} \rangle
 \langle  -\mathbf{q} - \frac{\mu_{\beta \gamma}}{m_{\beta}} \mathbf{q}' | t_{\beta \gamma}\left(-\frac{q'^2}{2 \mu_{\beta \gamma, \alpha}}\right) | \frac{\mu_{\beta \gamma}}{m_{\gamma}} \mathbf{q}' \rangle
\langle -\mathbf{q}' | t_{\gamma \alpha}(0) | \mathbf{0} \rangle \,d\mathbf{q}'.
\end{split}
\end{align}
%\end{widetext}
To analyze how Eqs.~\eqref{eq:TbG0TaG0Tb_momentum_space}--\eqref{eq:TgG0TaG0Tb_momentum_space} contribute to $\mathcal{U}_0$ near a $p$-wave dimer resonance, we set $p = 0$, so that the first and final $t$ operators only contribute via their $s$-wave components as a consequence of Eqs.~\eqref{eq:tlneq0_zero_p} and \eqref{eq:t0_small_p_pa_z}. To get the largest scaling in $a_{1,\beta \gamma}$, we consider the $p$-wave component of the second $t$ operator. We also take the limit $q \to 0$. We note that the $\sqrt{-a_{1,\beta \gamma}}$ behavior in Eq.~\eqref{eq:int_1/q^2_t1_dq_a1abs_to_inf} arises from an arbitrarily small integration interval. Therefore, we can take the first and final $t$ matrices outside the integrals in Eqs.~\eqref{eq:TbG0TaG0Tb_momentum_space}--\eqref{eq:TgG0TaG0Tb_momentum_space}. If we then multiply Eqs.~\eqref{eq:TbG0TaG0Tb_momentum_space}--\eqref{eq:TgG0TaG0Tb_momentum_space} by $1/\sqrt{-a_{1,\beta \gamma}}$ and take the limit $|a_{1,\beta \gamma}| \to \infty$ using Eq.~\eqref{eq:int_1/q^2_t1_dq_a1abs_to_inf}, we find
%\begin{widetext}
\begin{align}
\begin{split}\label{eq:TbG0TaG0Tb_p_0_pwave}
48 \pi &\, \mu_{\gamma \alpha}^2 \bigg(t_0^{(\gamma \alpha)}\left(0,0,0\right)\bigg)^2
\int_0^{\infty} \frac{1}{q'^2}
t_1^{(\beta \gamma)}\left(\frac{\mu_{\beta \gamma}}{m_{\gamma}} q', \frac{\mu_{\beta \gamma}}{m_{\gamma}} q',-\frac{q'^2}{2 \mu_{\beta \gamma,\alpha}}\right) \,dq' \frac{1}{\sqrt{-a_{1,\beta \gamma}}}
\\
&\underset{|a_1| \to \infty}{=} - \frac{3 \sqrt{2}}{8 \pi^4} \frac{\sqrt{\mu_{\beta \gamma,\alpha} \mu_{\beta \gamma}}}{m_{\gamma}^2 \hbar^4}
\frac{a_{\gamma \alpha}^2}{\sqrt{\tilde{r}_{1,\beta \gamma}}},
\end{split}
%%%%%%%%%%%%%%%%%%%%%%%%%%%%%%%%%%%%%%%%%%%%%%%%%%%
 \\
 \begin{split}\label{eq:TgG0TaG0Tg_p_0_pwave}
 48 \pi &\, \mu_{\alpha \beta}^2  \bigg(t_0^{(\alpha \beta)}\left(0,0,0\right)\bigg)^2
\int_0^{\infty} \frac{1}{q'^2}
t_1^{(\beta \gamma)}\left(\frac{\mu_{\beta \gamma}}{m_{\beta}} q', \frac{\mu_{\beta \gamma}}{m_{\beta}} q',-\frac{q'^2}{2 \mu_{\beta \gamma,\alpha}}\right) \,dq' 
\frac{1}{\sqrt{-a_{1,\beta \gamma}}}
\\
&\underset{|a_1| \to \infty}{=} - \frac{3 \sqrt{2}}{8 \pi^4} \frac{\sqrt{\mu_{\beta \gamma,\alpha} \mu_{\beta \gamma}}}{m_{\beta}^2 \hbar^4}
\frac{a_{\alpha \beta}^2}{\sqrt{\tilde{r}_{1,\beta \gamma}}},
\end{split}
 %%%%%%%%%%%%%%%%%%%%%%%%%%%%%%%%%%%%%%%%%%%%%%%%%%%
 \\
 \begin{split}\label{eq:TbG0TaG0Tg_p_0_pwave}
48 \pi &\, \mu_{\alpha \beta} \mu_{\gamma \alpha} 
 t_0^{(\gamma \alpha)}\left(0,0,0\right) t_0^{(\alpha \beta)}\left(0,0,0\right)
\int_0^{\infty} \frac{1}{q'^2}
t_1^{(\beta \gamma)}\left(\frac{\mu_{\beta \gamma}}{m_{\gamma}} q', -\frac{\mu_{\beta \gamma}}{m_{\beta}} q',-\frac{q'^2}{2 \mu_{\beta \gamma,\alpha}}\right) \,dq'
\frac{1}{\sqrt{-a_{1,\beta \gamma}}}
\\
&\underset{|a_1| \to \infty}{=} \frac{3 \sqrt{2}}{8 \pi^4} \frac{\sqrt{\mu_{\beta \gamma,\alpha} \mu_{\beta \gamma}}}{m_{\beta} m_{\gamma} \hbar^4}
\frac{a_{\gamma \alpha} a_{\alpha \beta}}{\sqrt{\tilde{r}_{1,\beta \gamma}}},
\end{split}
 %%%%%%%%%%%%%%%%%%%%%%%%%%%%%%%%%%%%%%%%%%%%%%%%%%%
 \\
 \begin{split}\label{eq:TgG0TaG0Tb_p_0_pwave}
 48 \pi &\, \mu_{\alpha \beta} \mu_{\gamma \alpha}  
 t_0^{(\gamma \alpha)}\left(0,0,0\right) t_0^{(\alpha \beta)}\left(0,0,0\right)
\int_0^{\infty} \frac{1}{q'^2}
t_1^{(\beta \gamma)}\left(-\frac{\mu_{\beta \gamma}}{m_{\beta}} q', \frac{\mu_{\beta \gamma}}{m_{\gamma}} q',-\frac{q'^2}{2 \mu_{\beta \gamma,\alpha}}\right) \,dq'
\frac{1}{\sqrt{-a_{1,\beta \gamma}}}
\\
&\underset{|a_1| \to \infty}{=} \frac{3 \sqrt{2}}{8 \pi^4} \frac{\sqrt{\mu_{\beta \gamma,\alpha} \mu_{\beta \gamma}}}{m_{\beta} m_{\gamma} \hbar^4}
\frac{a_{\gamma \alpha} a_{\alpha \beta}}{\sqrt{\tilde{r}_{1,\beta \gamma}}}.
 \end{split}
\end{align}
%\end{widetext}
Equations~\eqref{eq:TbG0TaG0Tb_p_0_pwave}--\eqref{eq:TgG0TaG0Tb_p_0_pwave} result in Eqs.~(7)--(9) of the main text.
\end{widetext}

\section{Three-body recombination into the shallow $p$-wave dimer state}

Here we determine the contribution to $\mathrm{Im} \left(\mathcal{U}_0 \right)$ that comes from three-body recombination into the shallow $p$-wave dimer state. Using the optical theorem, we will find that this contribution gives Eqs.~(7) and (9) of the main text in the limit $a_1 \to +\infty$.
%Here we derive Eq.~\eqref{eq:U0_limit_large_a1} in the limit $a_1 \to +\infty$ from the optical theorem given by Eq.~\eqref{eq:Im_U00_optical_theorem}.

The optical theorem for three-particle scattering connects the imaginary part of $\mathcal{U}_0$ to the three-body recombination rate \cite{schmid1974threebody}:
\begin{equation}\label{eq:Im_U00_optical_theorem}
\begin{aligned}
\frac{1}{(2 \pi)^6}\mathrm{Im} &\left(\mathcal{U}_0 \right)
=
-\pi \sum_{\substack{(\alpha, \beta, \gamma) = (1,2,3), \\ (2,3,1), \, (3,1,2)}}  \sum_{d} \mu_{\beta \gamma, \alpha} q_{\alpha, d}
\\
&\int 
\left\lvert{}_{\alpha}\langle \varphi_{d}^{(\beta \gamma)}, \mathbf{q}_{\alpha,d} |
U_{\alpha 0}(0) | \mathbf{0}, \mathbf{0} \rangle \right\rvert^2
\,d\hat{\mathbf{q}}_{\alpha,d}.
\end{aligned}
\end{equation}
Here $d$ labels the dimer states $\lvert \varphi_{d}^{(\beta \gamma)} \rangle$ consisting of particles $\beta$ and $\gamma$ whose bound state energy $E_{\mathrm{2b},d}$ fixes $q_{\alpha,d}$ via $E_{\mathrm{2b},d} = -q_{\alpha,d}^2/(2 \mu_{\beta \gamma, \alpha})$. These bound states are normalized as $\langle \varphi_{d}^{(\beta \gamma)} | \varphi_{d}^{(\beta \gamma)} \rangle = 1$. The label $d$ can be regarded as a collection of the quantum numbers $n$, $l$ and $m$ for which dimer states exist, i.e., $d = \lbrace n, l, m \rbrace$. Here $l$ and $m$ are the quantum numbers for the angular momentum and its projection on the quantization axis, respectively.
The dimer state $\lvert \varphi_{d}^{(\beta \gamma)} \rangle$ can be represented as
\begin{equation}\label{eq:varphi_nl_vs_g_nl}
\lvert \varphi_{d}^{(\beta \gamma)} \rangle = X_{n l}^{(\beta \gamma)} G_0^{(\mathrm{2b})}(E_{\mathrm{2b},d}) \lvert g_{n l m}^{(\beta \gamma)}(E_{\mathrm{2b},d}) \rangle,
\end{equation} 
where the Weinberg states $\lvert g_{n l m}^{(\beta \gamma)}\left(z_{\mathrm{2b}}\right) \rangle$ are defined as the eigenstates of $V_{\beta \gamma} G_0^{(\mathrm{2b})}(z_{\mathrm{2b}})$ \cite{weinberg1963expansion}. Here $X_{n l}^{(\beta \gamma)}$ is a normalization factor that is fixed via $\langle  \varphi_{d}^{(\beta \gamma)} |  \varphi_{d}^{(\beta \gamma)} \rangle = 1$ and can be chosen to be real. 

Our goal is to derive an expression for ${}_{\alpha}\langle \varphi_d^{(\beta \gamma)}, \mathbf{q}_{\alpha,d} | U_{\alpha 0}(0) | \mathbf{0}, \mathbf{0} \rangle$ for a weakly-bound $p$-wave dimer state. From Eq.~(1) of the main text, we find that
\begin{equation}
\begin{aligned}
U_{\alpha 0}(z) &= G_0^{-1}(z) + T_{\beta}(z) + T_{\gamma}(z) 
\\
&+ T_{\beta}(z) G_0(z) \left[T_{\gamma}(z) + T_{\alpha}(z)\right]
\\
&+ T_{\gamma}(z) G_0(z) \left[T_{\alpha}(z) + T_{\beta}(z)\right] + ...
\end{aligned}
\end{equation}
The first term gives ${}_{\alpha}\langle \varphi_d^{(\beta \gamma)}, \mathbf{q}_{\alpha,d} | G_0^{-1}(0) | \mathbf{0}, \mathbf{0} \rangle = 0$. In the limit of zero $p$-wave dimer binding energy, the dominant contribution to ${}_{\alpha}\langle \varphi_d^{(\beta \gamma)}, \mathbf{q}_{\alpha,d} | U_{\alpha 0}(0) | \mathbf{0}, \mathbf{0} \rangle$ comes from 
\begin{equation}
\begin{aligned}
{}_{\alpha}\langle \varphi_d^{(\beta \gamma)}, &\mathbf{q}_{\alpha,d}  | T_{\beta}(0) | \mathbf{0}, \mathbf{0} \rangle = -\frac{2 \mu_{\gamma \alpha}}{q_{\alpha,d}^2} \left(X_{n l}^{(\beta \gamma)}\right)^* 
\\
&
\langle g_{n l m}^{(\beta \gamma)}(E_{\mathrm{2b},d}) | \frac{\mu_{\beta \gamma}}{m_{\gamma}}\mathbf{q}_{\alpha,d} \rangle
\, \langle -\mathbf{q}_{\alpha,d} | t_{\gamma \alpha}(0) | \mathbf{0} \rangle
\end{aligned}
\end{equation}
and 
\begin{equation}
\begin{aligned}
{}_{\alpha}\langle \varphi_d^{(\beta \gamma)}, &\mathbf{q}_{\alpha,d} | T_{\gamma}(0) | \mathbf{0}, \mathbf{0} \rangle = -\frac{2 \mu_{\alpha \beta}}{q_{\alpha,d}^2} \left(X_{n l}^{(\beta \gamma)}\right)^* 
\\
&
\langle g_{n l m}^{(\beta \gamma)}(E_{\mathrm{2b},d}) | -\frac{\mu_{\beta \gamma}}{m_{\beta}}\mathbf{q}_{\alpha,d} \rangle
\, \langle \mathbf{q}_{\alpha,d} | t_{\alpha \beta}(0) | \mathbf{0} \rangle.
\end{aligned}
\end{equation}
Substituting these two terms for $l=1$ into Eq.~\eqref{eq:Im_U00_optical_theorem} and taking the limit $E_{\mathrm{2b},d} \to 0$, we find that they contribute to $\frac{1}{(2 \pi)^6}\mathrm{Im} \left(\mathcal{U}_0 \right)$ as
\begin{equation}\label{eq:ImU0_betagamma_a1pos}
\begin{aligned}
-\frac{3 \sqrt{2}}{8 \pi^4} 
\left(\frac{a_{\gamma \alpha}}{m_{\gamma}} - \frac{a_{\alpha \beta}}{m_{\beta}}\right)^2
\frac{\sqrt{\mu_{\beta \gamma,\alpha} \mu_{\beta \gamma}}}{\hbar^4 \sqrt{\tilde{r}_{1,\beta \gamma}}} \sqrt{a_{1,\beta \gamma}}.
\end{aligned}
\end{equation}
The factor $3$ comes from the degeneracy of the $p$-wave dimer state ($m = -1$, 0 or 1). In the derivation of Eq.~\eqref{eq:ImU0_betagamma_a1pos}, we have also used the following general property of the $p$-wave dimer state at $a_{1,\beta \gamma}\to \infty$:
\begin{equation}\label{eq:ReffPwave_vs_gn1}
\begin{aligned}
\left\lvert X_{n,1}^{(\beta \gamma)}\right\rvert^2 \left(\frac{\partial g_{n,1}^{(\beta \gamma)}\left(p,0\right)}{\partial p}\Bigg|_{p = 0}\right)^2 &=  \frac{1}{\pi} \frac{1}{\mu_{\beta \gamma}^2 \hbar \, \tilde{r}_{1,\beta \gamma}},
\end{aligned}
\end{equation}
where we defined the functions $g_{n l}^{(\beta \gamma)}\left(p,z_{\mathrm{2b}}\right)$ via $\langle \mathbf{p} | g_{n l m}^{(\beta \gamma)}\left(z_{\mathrm{2b}}\right) \rangle = Y_l^{m}(\hat{\mathbf{p}}) g_{n l}^{(\beta \gamma)}\left(p,z_{\mathrm{2b}}\right)$. Equation~\eqref{eq:ReffPwave_vs_gn1} can be derived from
\begin{equation}
\frac{\partial t_{\beta \gamma}(z_{\mathrm{2b}})}{\partial z_{\mathrm{2b}}} = - t_{\beta \gamma}(z_{\mathrm{2b}}) \left(G_0^{(\mathrm{2b})}(z_{\mathrm{2b}})\right)^2  t_{\beta \gamma}(z_{\mathrm{2b}})
\end{equation}
when evaluated at $z_{\mathrm{2b}} = 0$ in the limit $a_{1,\beta \gamma}\to \infty$. Equation~\eqref{eq:ImU0_betagamma_a1pos} is consistent with Eqs.~(7) and (9) of the main text. Note that the number of shallow $p$-wave dimer states consisting of dissimilar particles doubles for the BBX system compared to a system consisting of three dissimilar particles with one resonant $p$-wave interaction.

\section{Three-body transition amplitude of the BBX system}
\label{sec:U00_amplitude_BBX}

Here we analyze the three-body transition amplitude $\langle \mathbf{p}, \mathbf{q} | U_{00}(0) | \mathbf{0}, \mathbf{0} \rangle$ for a system consisting of two identical bosons (B) and one distinguishable particle (X) which is either bosonic or fermionic. It determines the zero-energy three-body scattering state by 
\begin{equation}
\lvert \Psi_{\mathrm{3b}}(0+ i 0)\rangle = \lvert \mathbf{0}, \mathbf{0} \rangle + G_0(0+ i 0)U_{00}(0+ i 0)\lvert \mathbf{0}, \mathbf{0} \rangle.
\end{equation}
Here $z = 0 + i 0$ means that we take the zero-energy limit from the upper half of the complex energy plane. In the following, we omit this complex part and write $z = 0$ for notational convenience. Our analysis of $\langle \mathbf{p}, \mathbf{q} | U_{00}(0) | \mathbf{0}, \mathbf{0} \rangle$ for the BBX system is similar to the analysis of Ref.~\cite{mestrom2019hypervolumeSqW} which applied to three identical bosons.

For the BBX system, we take particles $1$ and $2$ to be the identical bosons and particle $3$ to be X. Thus $T_{2}(z) =  P_{1,2} T_{1}(z) P_{1,2}$ where $P_{\alpha \beta}$ is the permutation operator for particles $\alpha$ and $\beta$. By defining $\breve{U}_{\alpha 0}(z) \equiv T_{\alpha}(z) G_0(z) U_{\alpha 0}(z) (1+P)$, where $P = P_{\alpha \beta} P_{\beta \gamma} + P_{\alpha \beta} P_{\alpha \gamma}$, Eq.~(1) of the main text transforms to
\begin{equation}\label{eq:AGS_BBX_with_BB}
\begin{aligned}
U_{0 0}(z)&(1+P)= (1+P_{1,2}) \breve{U}_{1, 0}(z) + \breve{U}_{3, 0}(z), \\
\breve{U}_{1, 0}(z) &= T_{1}(z)(1+P) 
\\
&+ T_{1}(z) G_0(z)  \left(P_{1,2} \breve{U}_{1, 0}(z) + \breve{U}_{3, 0}(z)\right),
\\
\breve{U}_{2, 0}(z) &= P_{1,2} \breve{U}_{1,0}(z),
\\
\breve{U}_{3,0}(z) &= T_{3}(z)(1+P) + T_{3}(z) G_0(z)  (1 + P_{1,2}) \breve{U}_{1,0}(z),
\end{aligned}
\end{equation}
and
\begin{equation}\label{eq:U00_vs_Ubreve_matrix}
\langle \mathbf{p}, \mathbf{q} | U_{00}(z) | \mathbf{0}, \mathbf{0} \rangle = \frac{1}{3} \sum_{\alpha = 1}^{3} \, {}_{\alpha}\langle \mathbf{p}_{\alpha}, \mathbf{q}_{\alpha}| \breve{U}_{\alpha 0}(z) | \mathbf{0}, \mathbf{0} \rangle.
\end{equation}
%The index $\xi$ in $\lvert \mathbf{p}_{\xi}, \mathbf{q}_{\xi} \rangle_{\xi}$ indicates the considered Jacobi momenta. So for $\xi = \alpha$, $\mathbf{p}_{\alpha}$ is the relative momentum between particles $\beta$ and $\gamma$, whereas $\mathbf{q}_{\alpha}$ is the relative momentum between particle $\alpha$ and the center of mass of the two-particle system $(\beta \gamma)$.

%\del{We note that the three-body operator $T_{\alpha}(z)$ is connected to the usual two-body transition operator $t_{\beta \gamma}(z_{\mathrm{2b}})$ that is defined via $t_{\beta \gamma}(z_{\mathrm{2b}}) = V_{\beta \gamma} + V_{\beta \gamma} \left(z_{\mathrm{2b}}-H_0^{(\mathrm{2b})}\right)^{-1} t_{\beta \gamma}(z_{\mathrm{2b}})$ \cite{taylor1972scattering}. Here $z_{\mathrm{2b}}$ is the two-body energy and $H_0^{(\mathrm{2b})}$ is the two-body kinetic energy operator in the two-body center-of-mass frame.}

We analyze the elements ${}_{\alpha}\langle \mathbf{p}_{\alpha}, \mathbf{q}_{\alpha}| \breve{U}_{\alpha 0}(0) | \mathbf{0}, \mathbf{0} \rangle$ by writing Eq.~\eqref{eq:AGS_BBX_with_BB} as an expansion,
\begin{equation}
\begin{aligned}
\breve{U}_{1, 0}(z) &= T_{1}(z)(1+P) + T_{1}(z) G_0(z) P_{1,2} T_{1}(z)(1+P)
\\
&+ T_{1}(z) G_0(z) T_{3}(z)(1+P) + ...
\\
\breve{U}_{3, 0}(z) &= T_{3}(z)(1+P) 
\\
&+ T_{3}(z) G_0(z)  (1 + P_{1,2}) T_{1}(z)(1+P) + ...,
\end{aligned}
\end{equation}
and analyzing each term. The first terms represent pure two-body scattering at zero energy:
\begin{equation}
\begin{aligned}
{}_{\alpha}\langle \mathbf{p}_{\alpha}, \mathbf{q}_{\alpha} | T_{\alpha}(0) | \mathbf{0}, \mathbf{0} \rangle
&= \delta(\mathbf{q}_{\alpha}) \langle \mathbf{p}_{\alpha} | t_{\beta \gamma}(0) | \mathbf{0} \rangle.
\end{aligned}
\end{equation}
Terms that involve two $T$ operators contribute to the $1/q_{\alpha}^2$ and $1/q_{\alpha}$ behavior of ${}_{\alpha}\langle \mathbf{0}, \mathbf{q}_{\alpha} | \breve{U}_{\alpha 0}(0) | \mathbf{0}, \mathbf{0} \rangle$. The $1/q_{\alpha}$ behavior is further determined by terms that involve three $T$ operators. The $\ln(q_{\alpha} \rho/\hbar)$ behavior arises from terms involving three and four $T$ operators. This analysis gives the following expressions for the coefficients $A_{\alpha}$, $B_{\alpha}$ and $C_{\alpha}$ in Eq.~(2) of the main text:
\begin{equation}
\begin{aligned}
A_{1} &= A_{2} = -\frac{1}{8 \pi^4} (1+ \chi) \frac{a_{\mathrm{BX}} \left(a_{\mathrm{BX}} + a_{\mathrm{BB}}\right)}{m_{\mathrm{X}} \hbar^2},
\end{aligned}
\end{equation}
\begin{equation}
\begin{aligned}
A_{3} &= -\frac{1}{2 \pi^4} \chi \frac{a_{\mathrm{BB}} a_{\mathrm{BX}}}{m_{\mathrm{X}} \hbar^2},
\end{aligned}
\end{equation}
\begin{widetext}
\begin{eqnarray}
\begin{aligned}
B_{1} &= B_{2} = \frac{1}{8 \pi^4} \left[(1+\chi)^2 \mathrm{arcsin}\left(\frac{1}{1+\chi}\right) 
- \sqrt{\chi(2+\chi)} \right]
\frac{\left(a_{\mathrm{BX}}\right)^2 \left(a_{\mathrm{BX}} + a_{\mathrm{BB}}\right)}{m_{\mathrm{X}} \hbar^3}
\\
&+ \frac{1}{2 \pi^4} (1+\chi) \mathrm{arcsin}\left(\frac{1}{2}\sqrt{\frac{2\chi}{1+ \chi}}\right) \frac{\left(a_{\mathrm{BX}}\right)^2 a_{\mathrm{BB}}}{m_{\mathrm{X}} \hbar^3},
\end{aligned}
\end{eqnarray}
\begin{eqnarray}
\begin{aligned}
B_{3} &= \frac{1}{2 \pi^4}(1+\chi)  \mathrm{arcsin}\left(\frac{1}{2}\sqrt{\frac{2\chi}{1+ \chi}}\right) \frac{a_{\mathrm{BX}} a_{\mathrm{BB}} \left(a_{\mathrm{BX}} + a_{\mathrm{BB}}\right)}{m_{\mathrm{X}} \hbar^3} 
- \frac{1}{4 \pi^4} \sqrt{\chi (2+\chi)} \frac{a_{\mathrm{BX}} \left(a_{\mathrm{BB}}\right)^2}{m_{\mathrm{X}} \hbar^3}, 
\end{aligned}
\end{eqnarray}
\begin{eqnarray}
\begin{aligned}
C_{1} &= C_{2} = \frac{1}{4 \pi^5} \left[(1+\chi)^2 \mathrm{arcsin}\left(\frac{1}{1+\chi}\right) 
- \sqrt{\chi(2+\chi)} \right] \frac{\left(a_{\mathrm{BX}}\right)^3 \left(a_{\mathrm{BX}} + a_{\mathrm{BB}}\right)}{m_{\mathrm{X}} \hbar^4}
\\
&+ \frac{1}{2 \pi^5} \left[ 2 (1+\chi)  \mathrm{arcsin}\left(\frac{1}{2}\sqrt{\frac{2\chi}{1+ \chi}}\right)
- \sqrt{\chi(2+\chi)} \right]
 \frac{\left(a_{\mathrm{BX}}\right)^2 \left(a_{\mathrm{BB}}\right)^2}{m_{\mathrm{X}} \hbar^4}
\\
&+ \frac{2}{\pi^5} (1+\chi) \mathrm{arcsin}\left(\frac{1}{2}\sqrt{\frac{2\chi}{1+ \chi}}\right) \frac{\left(a_{\mathrm{BX}}\right)^3 a_{\mathrm{BB}}}{m_{\mathrm{X}} \hbar^4},
\end{aligned}
\end{eqnarray}
and
\begin{eqnarray}
\begin{aligned}
C_{3} &=  \frac{1}{2 \pi^5} \left[(1+\chi)^2 \mathrm{arcsin}\left(\frac{1}{1+\chi}\right) - \sqrt{\chi(2+\chi)}\right] \frac{\left(a_{\mathrm{BX}}\right)^2 a_{\mathrm{BB}}\left(a_{\mathrm{BX}}+ a_{\mathrm{BB}}\right)}{m_{\mathrm{X}} \hbar^4} 
\\
&+ \frac{2}{\pi^5} (1+\chi) \mathrm{arcsin}\left(\frac{1}{2}\sqrt{\frac{2\chi}{1+ \chi}}\right) \frac{\left(a_{\mathrm{BX}}\right)^2 \left(a_{\mathrm{BB}}\right)^2}{m_{\mathrm{X}} \hbar^4}. 
\end{aligned}
\end{eqnarray}
\end{widetext}

%-----------------

%There are three terms that involve two $T$ operators:
%\begin{equation}\label{eq:TG0PabT1+P_momentum_space}
%\begin{aligned}
%{}_{\alpha} &\langle \mathbf{p}, \mathbf{q} | T_{\alpha}(0) G_0(0) P_{\alpha \beta} T_{\alpha}(0)| \mathbf{0}, \mathbf{0} \rangle
%\\&= -\frac{1}{8 \pi^4} (1+ \chi) \frac{(a^{(\alpha)})^2}{m_{\gamma} \hbar^2} \frac{1}{q^2} 
%\\
%&- \frac{1}{8 \pi^4} \sqrt{\chi(2+\chi)} \frac{(a^{(\alpha)})^3}{m_{\gamma} \hbar^3} \frac{1}{q} + O\left(q^0\right).
%\end{aligned}
%\end{equation}
%\del{Plan = do not write down every expansion term. That is too much work!}

\section{Integral equations for the BBX system with zero BB interaction}

In this section, we write out the integral equations that determine ${}_{\alpha}\langle \mathbf{p}, \mathbf{q} | \breve{U}_{\alpha 0}(0) | \mathbf{0}, \mathbf{0} \rangle$ for the BBX system. We set the BB interaction to zero, so that ${}_{3}\langle \mathbf{p}, \mathbf{q} | \breve{U}_{3, 0}(0) | \mathbf{0}, \mathbf{0} \rangle = 0$. Here particle 3 is again chosen to be X. Equation \eqref{eq:AGS_BBX_with_BB} thus simplifies to
\begin{equation}\label{eq:AGS_BBX_with_no_BB}
\begin{aligned}
& U_{0 0}(z)(1+P)= (1+P_{1,2}) \breve{U}_{1,0}(z), \\
& \breve{U}_{1,0}(z) = T_{1}(z)(1+P) + T_{1}(z) G_0(z) P_{1,2} \breve{U}_{1,0}(z).
\end{aligned}
\end{equation}
Next, we expand ${}_{1}\langle \mathbf{p}, \mathbf{q} | \breve{U}_{1, 0}(0) | \mathbf{0}, \mathbf{0} \rangle$ as
\begin{equation}\label{eq:breve_Ualpha0_vs_breve_A_nl}
\begin{aligned}
{}_{1}\langle \mathbf{p}, \mathbf{q} | &\breve{U}_{1, 0}(0) | \mathbf{0}, \mathbf{0} \rangle
= 3 \, \langle \mathbf{p} | t_{\mathrm{BX}}(0) | \mathbf{0} \rangle \delta(\mathbf{q})
\\
&+
3 \sum_{l = 0}^{\infty} \frac{(-1)^l}{\sqrt{2 l + 1}}  
\sum_{m  = -l}^{l} 4 \pi \,Y_l^{m}(\hat{\mathbf{p}}) \left[ Y_{l}^{m}(\hat{\mathbf{q}})\right]^*
\\
&\sum_{n = 1}^{\infty} \tau_{n l}\left(-\frac{q^2}{2 \mu_{\mathrm{BX, B}}}\right) g_{n l}\left(p, -\frac{q^2}{2 \mu_{\mathrm{BX, B}}}\right) \breve{A}_{n l}(q).
\end{aligned}
\end{equation}
Here we have defined some new functions $\tau_{n l}\left(z_{\mathrm{2b}}\right)$, $g_{n l}\left(p,z_{\mathrm{2b}}\right)$ and $\breve{A}_{n l}(q)$. The functions $\tau_{n l}\left(z_{\mathrm{2b}}\right)$ and $g_{n l}\left(p,z_{\mathrm{2b}}\right)$ are defined via the Weinberg expansion of $t_{\mathrm{BX}}\left(z_{\mathrm{2b}}\right)$ \cite{weinberg1963expansion},
\begin{equation}
\begin{aligned}
t_{\mathrm{BX}}(z_{\mathrm{2b}}) &= -4 \pi \sum_{l=0}^{\infty} \sum_{n = 1}^{\infty} \tau_{n l}\left(z_{\mathrm{2b}}\right) 
\\
&\phantom{=} \times \sum_{m = -l}^{l}  \lvert g_{n l m}\left(z_{\mathrm{2b}}\right) \rangle \langle g_{n l m}\left(z_{\mathrm{2b}}\right) \rvert,
\end{aligned}
\end{equation}
and $\langle \mathbf{p} | g_{n l m}\left(z_{\mathrm{2b}}\right) \rangle = Y_l^{m}(\hat{\mathbf{p}}) g_{n l}\left(p,z_{\mathrm{2b}}\right)$. The Weinberg states $\lvert g_{n l m}\left(z_{\mathrm{2b}}\right) \rangle$ are eigenstates of $V_{\mathrm{BX}} G_0^{(\mathrm{2b})}(z_{\mathrm{2b}})$. The corresponding eigenvalues determine $\tau_{n l}\left(z_{\mathrm{2b}}\right)$. We use the same definitions for $\tau_{n l}\left(z_{\mathrm{2b}}\right)$ and $g_{n l}\left(p,z_{\mathrm{2b}}\right)$ as those presented in section IV B of Ref.~\cite{mestrom2019squarewell}.
%We also define the functions $g_{n l}\left(p,z_{\mathrm{2b}}\right)$ by $\langle \mathbf{p} | g_{n l m}\left(z_{\mathrm{2b}}\right) \rangle \equiv Y_l^{m}(\hat{\mathbf{p}}) g_{n l}\left(p,z_{\mathrm{2b}}\right)$. 

%\begin{equation}
%\langle \mathbf{p} | T_{\mathrm{BX}}(z_{\mathrm{2b}}) | \mathbf{p}' \rangle = \sum_{l=0}^{\infty} (2 l + 1) P_l\left(\hat{\mathbf{p}}\cdot \hat{\mathbf{p}}'\right) t_l\left(p,p',z_{\mathrm{2b}}\right),
%\end{equation}
%where
%\begin{equation}
%t_l\left(p,p',z_{\mathrm{2b}}\right) = - \sum_{n = 1}^{\infty} \tau_{n l}\left(z_{\mathrm{2b}}\right) g_{n l}\left(p,z_{\mathrm{2b}}\right) g_{n l}\left(p',z_{\mathrm{2b}}\right).
%\end{equation}
%The functions $P_l(x)$ represent the Legendre polynomials.
%We use the same definitions for $\tau_{n l}\left(z_{\mathrm{2b}}\right)$ and $g_{n l}\left(p,z_{\mathrm{2b}}\right)$ as in section IV B of Ref.~\cite{mestrom2019squarewell}.

From Eq.~\eqref{eq:AGS_BBX_with_no_BB} and \eqref{eq:breve_Ualpha0_vs_breve_A_nl} we derive the following integral equation for $\breve{A}_{n l}(q)$:
\begin{equation}\label{eq:breve_mathsfA_nl_q_BBX}
\begin{aligned}
\breve{A}_{n l}(q) =& - \sum_{n' = 1}^{\infty} \tau_{n',0}(0) g_{n',0}(0,0) U_{n l, n', 0}(q,0)
\\
&+ 4 \pi \sum_{n' l'} \int_0^{\infty} \tau_{n' l'}\left(-\frac{q'^2}{2 \mu_{\mathrm{BX,B}}}+ i 0\right)
\\
& U_{n l, n' l'}(q,q') \breve{A}_{n' l'}(q') \,q'^2 \,dq',
\end{aligned}
\end{equation}
where
\begin{widetext}
\begin{equation} \label{eq:U_nln1l1^alpha_alpha_BBX}
\begin{aligned}
U_{n l,n' l'}(q,q') &= \frac{2 m_{\mathrm{X}}}{1 + \chi} \frac{1}{4 \pi} (-1)^{l+l'}
\sqrt{2 l + 1} \sqrt{2 l' + 1}
\int P_{l}(\hat{\mathbf{q}} \cdot \reallywidehat{\mathbf{q}' + \frac{1}{1+\chi} \mathbf{q}})
P_{l'}(\reallywidehat{\mathbf{q} + \frac{1}{1+\chi} \mathbf{q}'} \cdot \hat{\mathbf{q}}')
\frac{1}{q^2 + q'^2 + \frac{2}{1 + \chi} \mathbf{q}\cdot \mathbf{q}'}
\\
&\phantom{=} g_{n l}\left(|\mathbf{q}' + \frac{1}{1+\chi} \mathbf{q}|,-\frac{q^2}{2 \mu_{\mathrm{BX,B}}}\right)
g_{n' l'}\left(|\mathbf{q} + \frac{1}{1+\chi} \mathbf{q}'|,-\frac{q'^2}{2 \mu_{\mathrm{BX,B}}}\right)
\,d\hat{\mathbf{q}}'.
\end{aligned}
\end{equation}
\end{widetext}
The functions $P_l(x)$ represent the Legendre polynomials.
The real part of $\breve{A}_{n,0}(q)$ contains singular terms that are propertional to $1/q^2$, $1/q$ and $\ln(q \rho/\hbar)$. Here $\rho$ is an arbitrary length scale. These singular terms can be derived from Eq.~\eqref{eq:breve_mathsfA_nl_q_BBX} by iteration and their prefactors are consistent with the results of section~\ref{sec:U00_amplitude_BBX}. Once they are known, they can be subtracted from $\breve{A}_{n,0}(q)$ before solving Eq.~\eqref{eq:breve_mathsfA_nl_q_BBX}. We solve the resulting integral equation by discretizing the momentum $q$. From the solution, we extract $\mathcal{U}_0$ via Eq.~\eqref{eq:U00_vs_Ubreve_matrix} and Eqs.~(2) and (3) of the main text.
 %The solution to the resulting integral equation is nonsingular in $q$ for $l = 0$.
%\end{widetext}

\section{$S$-wave resonance}

In this section we analyze $\mathcal{U}_{0}$ for the BBX system near an $s$-wave BX dimer resonance. We note that we take $\rho = |a_{\mathrm{BX}}|$ in Eq.~(2) of the main text, which uniquely defines $\mathcal{U}_{0}$. For resonant $s$-wave interactions ($|a_{\mathrm{BX}}| \to \infty$), the Efimov effect \cite{efimov1970energy, efimov1971weakly, efimov1973energy, braaten2006universality, naidon2017review, dincao2018review, helfrich2010heteronuclearBBX, mikkelsen2015K3BBX} causes $\mathcal{U}_{0}$ to be a log-periodic function of $a_{\mathrm{BX}}$. The corresponding universal expressions are given by
\begin{widetext}
\begin{equation}\label{eq:U0_large_a0neg}
\begin{aligned}
\frac{\mathcal{U}_0 m_{\mathrm{X}} \hbar^4}{C a_{\mathrm{BX}}^4} &\approx c_- + \frac{1}{2} b \frac{\sin\Big(2 s_0 \, \text{ln}(a_{\mathrm{BX}}/a_-) \Big) - i \sinh(2 \eta)}{\sin^2 \Big( s_0 \, \text{ln}(a_{\mathrm{BX}}/a_-) \Big) + \sinh^2(\eta)}
\end{aligned}
\end{equation}
%\begin{equation}\label{eq:U0_large_a0neg}
%\begin{aligned}
%\mathcal{U}_0 & m_{\mathrm{X}} \hbar^4/a_{\mathrm{BX}}^4 \approx C \Bigg(c_- + b \cot \Big(s_0 \, \text{ln}(a_{\mathrm{BX}}/a_-) + i \eta \Big)\Bigg)
%\\
%&= C \Bigg(c_- + \frac{1}{2} b \frac{\sin\Big(2 s_0 \, \text{ln}(a_{\mathrm{BX}}/a_-) \Big) - i \sinh(2 \eta)}{\sin^2 \Big( s_0 \, \text{ln}(a_{\mathrm{BX}}/a_-) \Big) + \sinh^2(\eta)} \Bigg).
%\end{aligned}
%\end{equation}
for $a_{\mathrm{BX}}<0$ and
%\begin{widetext}
\begin{eqnarray}\label{eq:U0_large_a0pos}
\begin{aligned}
\frac{\mathcal{U}_0 m_{\mathrm{X}} \hbar^4}{C a_{\mathrm{BX}}^4} &\approx  c_+ - \frac{\frac{1}{2} b \sin\left(2 s_0 \ln(a_{\mathrm{BX}}/a_+)\right)
+ i \left(\pi \sin^2\left(s_0 \ln(a_{\mathrm{BX}}/a_+)\right) + \pi \sinh^2(\eta) +\frac{1}{2} b \sinh(2\eta)\right)}{\cos^2\left( s_0 \ln(a_{\mathrm{BX}}/a_+)\right)+ \sinh^2(\pi s_0 + \eta)}
\end{aligned}
\end{eqnarray}
\end{widetext}
for $a_{\mathrm{BX}}>0$ \cite{noteU0Efimov}. Here $b = \pi \coth(\pi s_0)$ and $C = 32 \pi \left[(1+\chi)^2 \mathrm{arcsin}\left(\frac{1}{1+\chi}\right) - \sqrt{\chi(2+\chi)}\right]$. The coefficient $s_0$ sets the scaling factor $e^{\pi/s_0}$ of the Efimov effect and is a function of $\chi$ \cite{helfrich2010heteronuclearBBX}. 
%and is determined by
%\begin{equation}
%s_0 \cosh\left(\frac{\pi s_0}{2}\right) - \frac{2 \sinh\left[s_0 \arcsin\left(\frac{1}{1+\chi}\right)\right]}{\sin\left[2 \arcsin\left(\frac{1}{1+\chi}\right)\right]}= 0.
%\end{equation}
For $\chi \to \infty$, $s_0$ vanishes as $4/(\sqrt{3} \pi \chi)$, whereas $s_0$ diverges as $0.401031/\sqrt{\chi}$ for $\chi \to 0$. %Thus the Efimov period $e^{\pi/s_0}$ increases strongly for increasing $\chi$. 
%For example, $\chi = 0.5$, 1 and 2 give $e^{\pi/s_0} \approx 153.8$, $1986$ and $2.016\cdot 10^5$, respectively. Therefore, small mass ratios are most useful for the experimental observation of the Efimov effect. In the following, we consider $\chi \leq 0.5$, so that $s0 \geq 0.6238$.
Equations~\eqref{eq:U0_large_a0neg} and \eqref{eq:U0_large_a0pos} depend on the nonuniversal parameters $a_-$, $a_+$ and $\eta$ that are determined by the specific interparticle interactions. The coefficient $\eta$ sets the loss rate to deep dimer states \cite{braaten2006universality, dincao2018review}, whereas $a_{\pm}$ fix the Efimov spectrum. We have derived Eqs.~\eqref{eq:U0_large_a0neg} and \eqref{eq:U0_large_a0pos} using the analytical formulas for $\mathrm{Im}\left(\mathcal{U}_{0}\right)$ presented in Ref.~\cite{helfrich2010heteronuclearBBX} and the fact that the effects of deep dimer states on resonance can be deduced by substituting $s_0 \, \mathrm{ln}(a_{\mathrm{BX}}/a_{\pm}) \to s_0 \, \mathrm{ln}(a_{\mathrm{BX}}/a_{\pm}) + i \eta$ \cite{braaten2006universality}.
Equations~(\ref{eq:U0_large_a0neg}) and (\ref{eq:U0_large_a0pos}) complete the expressions for $\mathrm{Re}\left(\mathcal{U}_0\right)$ presented in Refs.~\cite{efimov1979threebody, braaten2002diluteBEC, braaten2006universality, dincao2018review, mestrom2019hypervolumeSqW}. The coefficients $c_{\pm}$ depend on $\chi$ and have not been previously calculated for the BBX system. 
% See Mathematica Week 44, Sunday 14 June 2020: this script checks the universal limits of U0 for a0<0 and a0>0.

To study $\mathcal{U}_0$ numerically at large $|a_{\mathrm{BX}}|$, we use a contact interaction as BX interaction and we set the BB interaction to zero. To fix the Efimov spectrum, we add a momentum cutoff $\Lambda$ to the contact interaction, i.e.,
\begin{equation} \label{eq:V_sepCut}
V_{\mathrm{BX}} = - \zeta \lvert g \rangle \langle g \rvert,
\end{equation}
where
\begin{equation}
\langle \mathbf{p} | g \rangle = 
\begin{cases}1,&\mbox{$0\leq p \leq \Lambda$},\\0,&\mbox{$p> \Lambda$}.\end{cases}
\end{equation}
We tune the scattering length $a_{\mathrm{BX}}$ by varying the interaction strength $\zeta$. For the separable potential in Eq.~\eqref{eq:V_sepCut}, there is only one $s$-wave dimer resonance. It does not support any other dimer states, so that $\eta = 0$ in Eqs.~(\ref{eq:U0_large_a0neg}) and (\ref{eq:U0_large_a0pos}).

\begin{figure}[btp] %[hb!]
    \centering
    \includegraphics[width=3.4in]{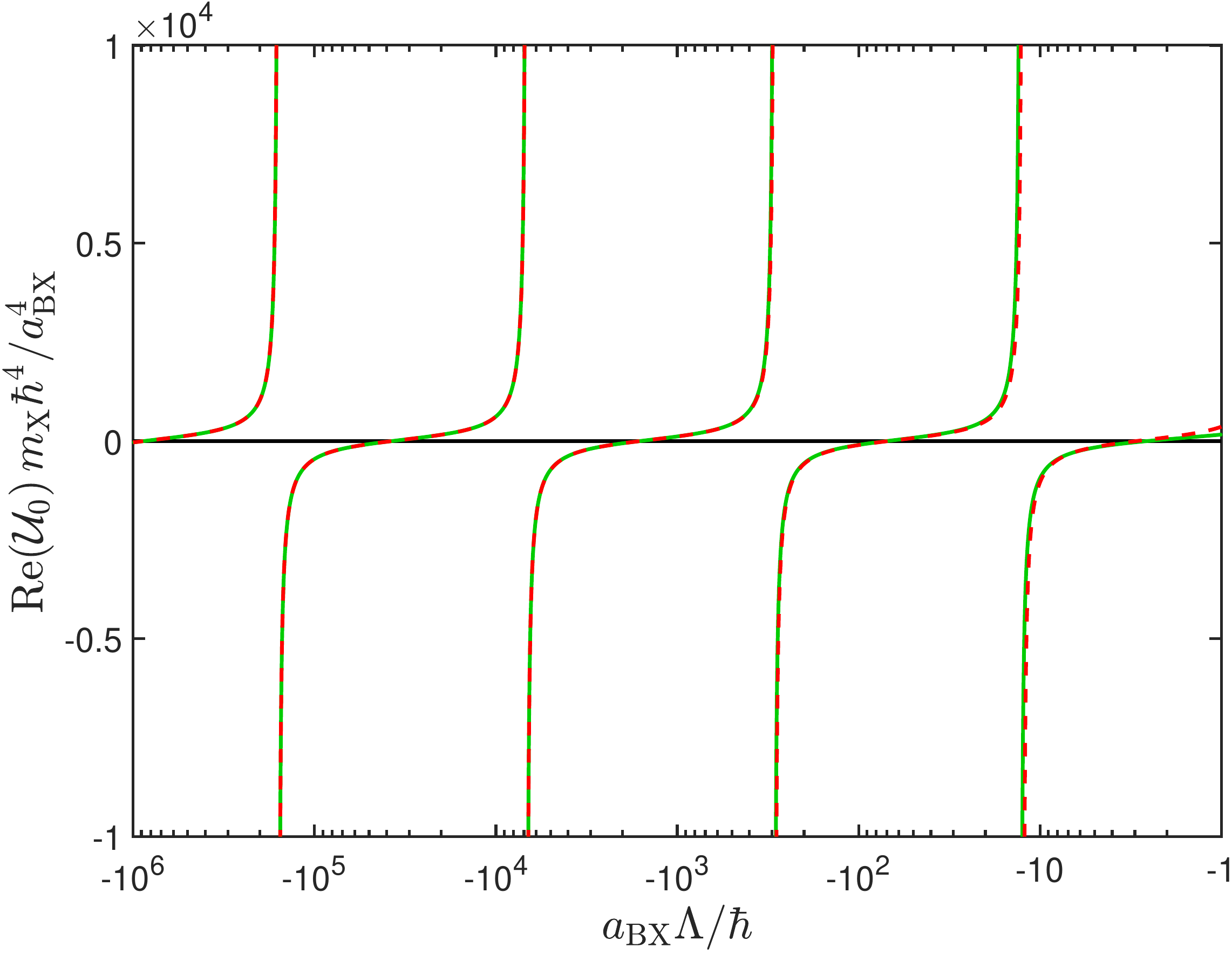}
    \caption{$\mathcal{U}_0$ (green solid line) near the BX dimer resonance of the contact interaction with momentum cutoff $\Lambda$ for $\chi = 0.2$ at $a_{\mathrm{BX}}<0$. The BB interaction is set to zero. The dashed curve represents the analytic zero-range result given by Eq.~(\ref{eq:U0_large_a0neg}) where we set $a_- \Lambda/\hbar = -1.583 \cdot 10^5$, $c_- = 0.40474$ and $\eta = 0$.}
    \label{fig:sepCut_U0_BBX_chi_0d2_a0neg_-1_-1e6}
\end{figure}

\begin{figure}[btp]
    \centering
    \includegraphics[width=3.4in]{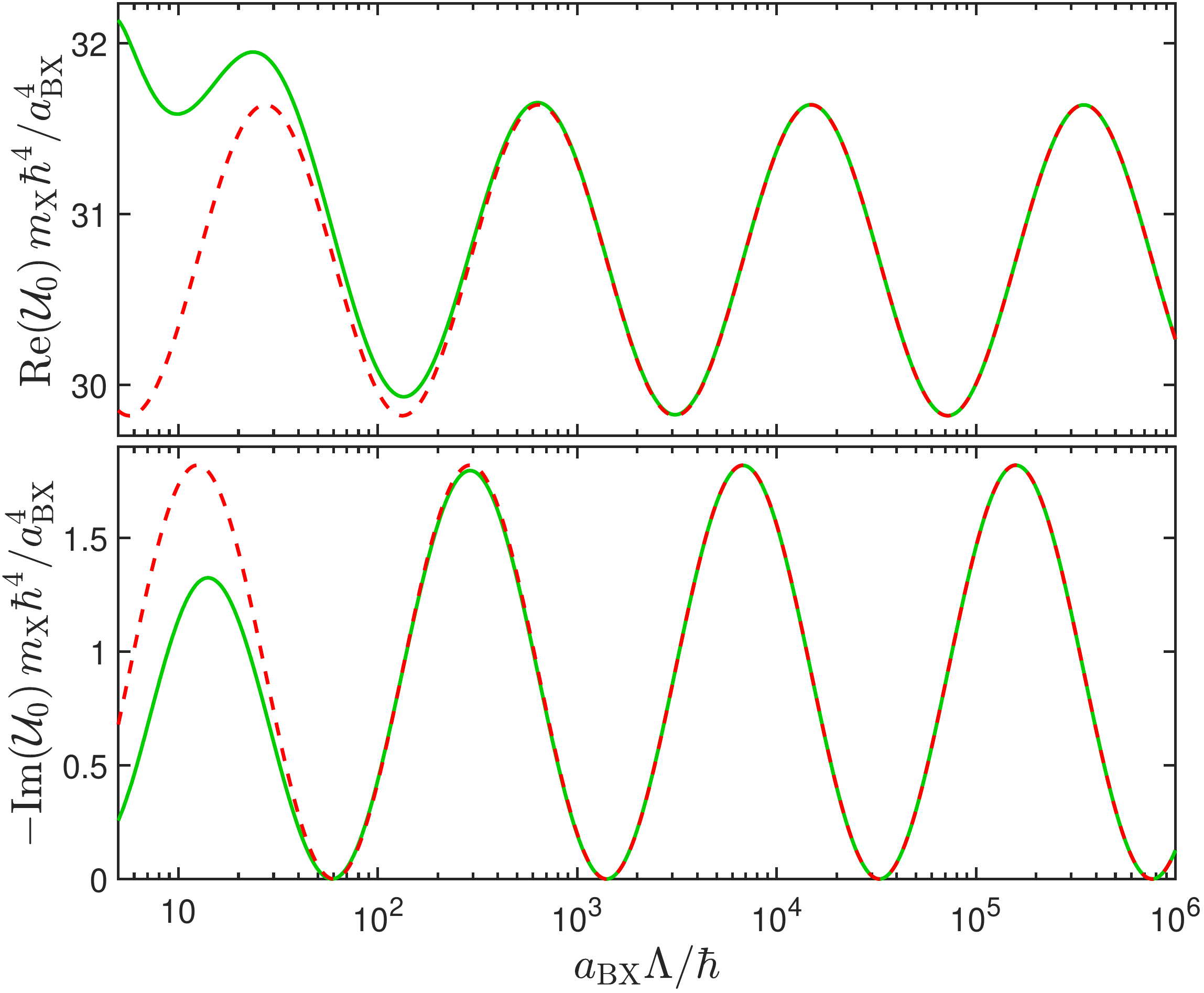}
    \caption{$\mathcal{U}_0$ (green solid line) near the BX dimer resonance of the contact interaction with momentum cutoff $\Lambda$ for $\chi = 0.2$ at $a_{\mathrm{BX}}>0$. The BB interaction is set to zero. The dashed curves represent the analytic zero-range result given by Eq.~(\ref{eq:U0_large_a0pos}) where we set $a_+ \Lambda/\hbar = 7.646 \cdot 10^5$, $c_+ = 0.40474$ and $\eta = 0$.} 
    \label{fig:sepCut_U0_BBX_chi_0d2_a0pos_1_1e6}
\end{figure}

\begin{figure}[btp]
%	\begin{subfigure}
%	\centering
%	\includegraphics[width=3.4in]{Figure_BBX_2Div3_C_bPlusMinus_vs_chi_v1.pdf}
%	\end{subfigure}
%\quad
%	\begin{subfigure}
	\centering
	\includegraphics[width=3.4in]{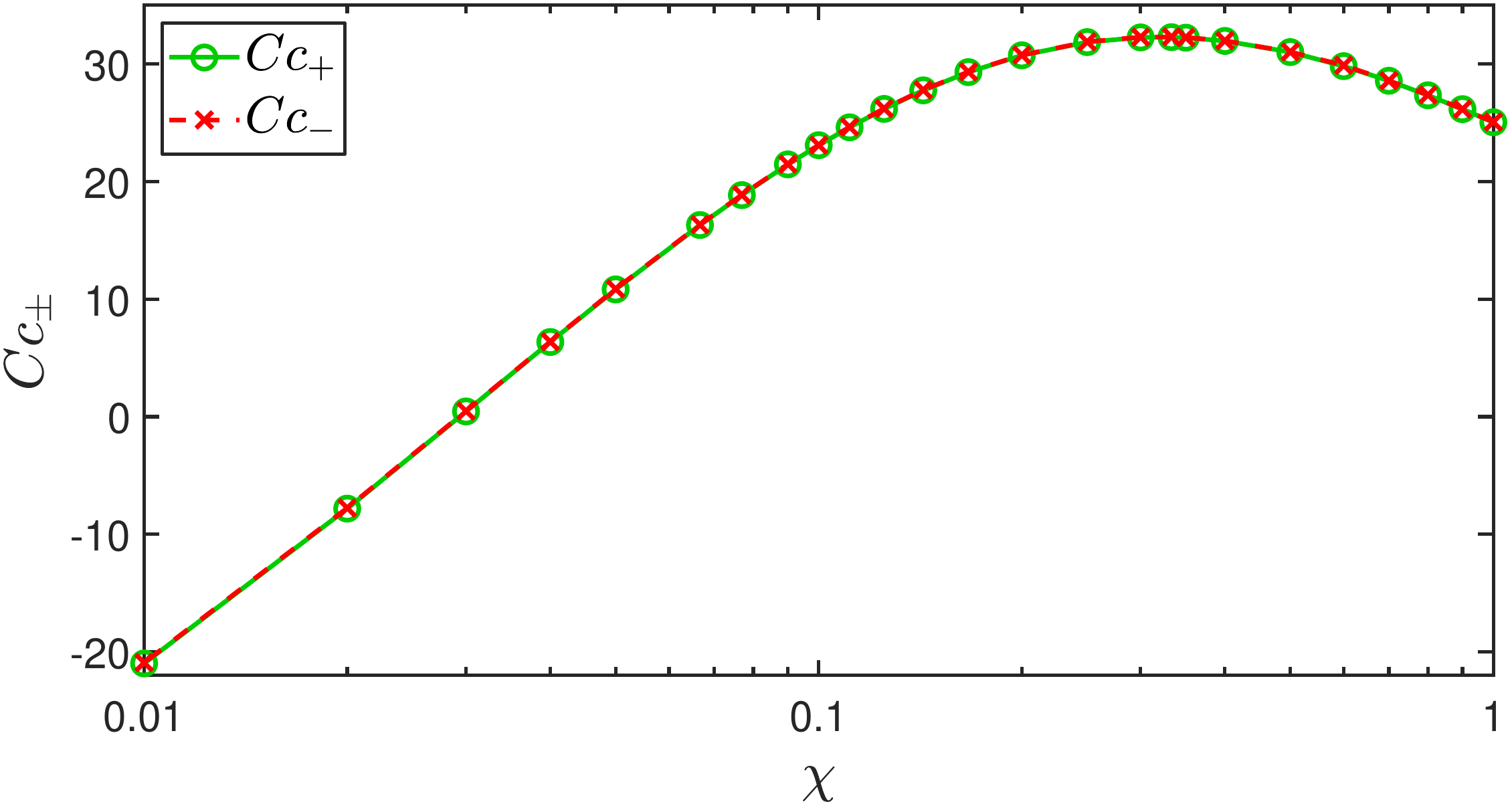}
%	\end{subfigure}
    \caption{The coeffients $C c_{\pm}$ as a function of the mass ratio $\chi$. The circles and crosses display our numerical results.}
    \label{fig:cPlusMinus_vs_chi_BBX}
\end{figure}

Figures~\ref{fig:sepCut_U0_BBX_chi_0d2_a0neg_-1_-1e6} and \ref{fig:sepCut_U0_BBX_chi_0d2_a0pos_1_1e6} show the behavior of $\mathcal{U}_0$ for $\chi = 0.2$ at $a_{\mathrm{BX}}<0$ and $a_{\mathrm{BX}}>0$, respectively. At large $|a_{\mathrm{BX}}|$, our results match the universal limits of Eq.~(\ref{eq:U0_large_a0neg}) and (\ref{eq:U0_large_a0pos}) from which we determine the coefficients $c_{\pm}$ for $\chi = 0.01$ to $1$. For larger values of $\chi$, $c_{\pm}$ are harder to determine numerically due to the increasing Efimov period $e^{\pi/s_0}$. Our results for $c_{\pm}$ are shown in Fig.~\ref{fig:cPlusMinus_vs_chi_BBX} and Table~\ref{tab:BBX_sepCut_chi_c+_c-}. For all values of $\chi$, we find that $c_+ = c_-$ within our numerical accuracy. This equivalency originates from the fact that both $c_+$ and $c_-$ represent hard-hypersherelike collisions at a hyperradius $|a_{\mathrm{BX}}|$ \cite{dincao2018review}. The prefactors $C c_{\pm}$ are much smaller compared to the value $1689$ that was found for three identical bosons (BBB) \cite{mestrom2019hypervolumeSqW}. This reduction in hard-hypersherelike scattering is due to the absence of one resonant interaction in the BBX system compared to the BBB sytem.

% Figure made with the code Plot_and_fit_Elastic_K3_List_BBX_v15_sepCut_chi_0d2_a0neg_NoFit.m

% Figure made with the code Plot_and_fit_Elastic_K3_List_BBX_v15_sepCut_chi_0d2_a0pos_NoFit.m

% Figure made with code Plot_BBX_universal_coefficients_v3.m

\begin{table}[btp]%[h!]%[htb!]
  \centering
  \caption{Values of the coefficients $c_{\pm}$ for the BBX system with various mass ratios $\chi$. These values are determined using the same approach as presented in the Supplemental Material of Ref.~\cite{mestrom2019hypervolumeSqW} for the BBB system. We have used the BX interaction given in Eq.~\eqref{eq:V_sepCut} and have set the BB interaction to zero.}
  \label{tab:BBX_sepCut_chi_c+_c-}
    \begin{ruledtabular}
    \begin{tabular}{cccc}
    \toprule
    $\chi$ & $e^{\pi/s_0}$ & $c_+$ & $c_-$ \\
    \hline
    $1$ & 1986 & 0.6873(5) & 0.688(2) \\ 
    $0.5$ & 153.8 & 0.5887(4) & 0.589(2) \\ 
    $0.2$ & 23.33 & 0.40474(5) & 0.4050(6)\\
    $0.1$ & 9.758 & 0.24894(5) & 0.2490(3) \\
    $0.05$ & 5.253 & 0.10059(5) & 0.101(1)\\
    $0.02$ & 2.947 & -0.0634(1) &  -0.063(1) \\   
    $0.01$ & 2.168 & -0.1585(2) &  -0.158(1) \\
    \bottomrule
    \end{tabular}\\
    \end{ruledtabular}
\end{table}

\begin{figure}[btp]%[btp] %[h!]%[hbtp]
	\centering
	\includegraphics[width=3.4in]{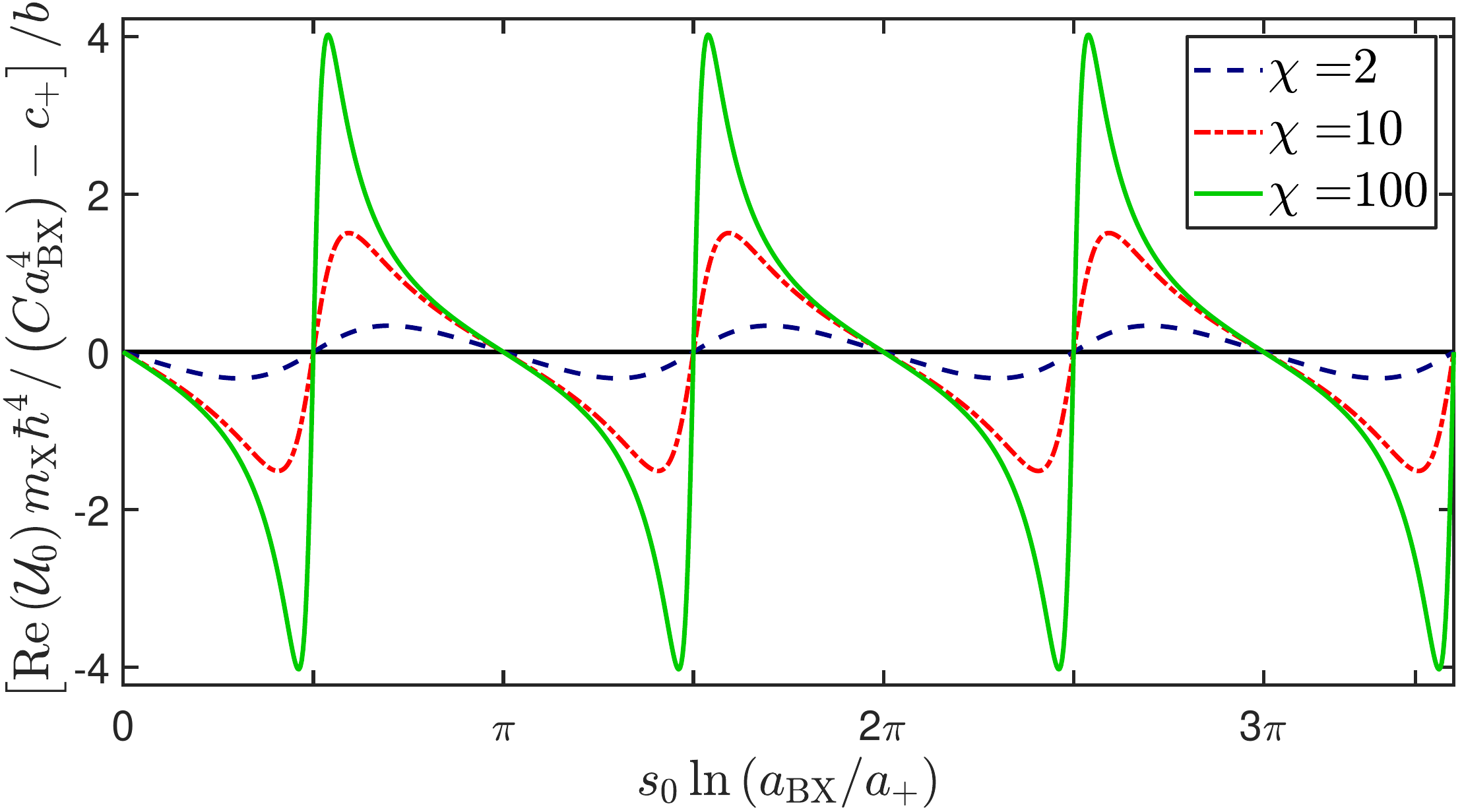}
%	\end{subfigure}
    \caption{Log-periodic behavior of $\mathrm{Re}\left(\mathcal{U}_0\right)$ near an $s$-wave BX dimer resonance for various mass ratios at $a_{\mathrm{BX}}>0$ as presented by Eq.~\eqref{eq:U0_large_a0pos}. We set $\eta = 0.1$.}
    \label{fig:ReU0_BBX_chi_2_10_100_a0pos}
\end{figure}

Equation (\ref{eq:U0_large_a0pos}) also demonstrates the existence of Efimov resonances at $a_{\mathrm{BX}}>0$ for large $\chi$ \cite{helfrich2010heteronuclearBBX}. By comparing Eqs.~(\ref{eq:U0_large_a0neg}) and (\ref{eq:U0_large_a0pos}), we find that $\mathrm{Re}\left(\mathcal{U}_0\right)$ behaves very similar for $a_{\mathrm{BX}}<0$ and $a_{\mathrm{BX}}>0$ when $\pi s_0\lesssim \eta$. Therefore, Efimov resonances do not only show up at $a_{\mathrm{BX}}<0$, but also at $a_{\mathrm{BX}}>0$ for $\chi \gg 1$ and $\eta \ll 1$ as illustrated in Fig.~\ref{fig:ReU0_BBX_chi_2_10_100_a0pos}. %in which case $\pi s_0$ determines the loss rate into the weakly-bound $s$-wave dimer state. 
At $a_{\mathrm{BX}}>0$, the specific behavior of $\mathrm{Re}\left(\mathcal{U}_0\right)$ for $\chi \gg 1$ suggests that the energies of the three-body quasibound states corresponding to these Efimov resonances increase for decreasing $a_{\mathrm{BX}}$. This is consistent with the conclusions of Ref.~\cite{helfrich2010heteronuclearBBX}. By analyzing atom-dimer scattering, Ref.~\cite{helfrich2010heteronuclearBBX} conjectured that the trimer resonances above the atom-dimer threshold originate from Efimov states that cross this threshold.

\end{document}